\documentclass[pra,twocolumn,notitlepage,showpacs]{revtex4-1}
\usepackage{graphicx}
\usepackage{amsmath}
\usepackage{amsfonts}
\usepackage{amssymb}
\usepackage{times}
\usepackage{amsfonts,amsmath,amssymb}
\usepackage{epsfig}
\usepackage{color}
\usepackage{epstopdf}
\DeclareGraphicsExtensions{.pdf,.eps,.png,.jpg,.mps}

\begin{document}

\title{Time-domain Ramsey interferometry with interacting Rydberg atoms}

\author{Christian Sommer$^{1,2,*}$, Guido Pupillo$^{3}$, Nobuyuki Takei$^{1,2}$, Shuntaro Takeda$^{1,2}$, Akira Tanaka$^{1}$, Kenji Ohmori$^{1,2}$, and Claudiu Genes$^{4,5}$}
\email{sommer@ims.ac.jp}

\affiliation{$^{1}$Institute for Molecular Science, National Institutes of Natural Sciences, Myodaiji, Okazaki 444-8585, Japan\\
$^{2}$SOKENDAI (The Graduate University for Advanced Studies), Myodaiji, Okazaki 444-8585, Japan\\
$^{3}$IPCMS (UMR 7504) and ISIS (UMR 7006), University of Strasbourg and CNRS, 67000 Strasbourg, France\\
$^{4}$Institut f\"ur Theoretische Physik, Universit\"at Innsbruck, Technikerstrasse 25, A-6020 Innsbruck, Austria\\
$^{5}$Vienna Center for Quantum Science and Technology, TU Wien-Atominstitut, Stadionallee 2, 1020 Vienna, Austria}

\date{\today}

\begin{abstract}
We theoretically investigate the dynamics of a gas of strongly interacting Rydberg atoms subject to a time-domain Ramsey interferometry protocol. The many-body dynamics is governed by an Ising-type Hamiltonian with long-range interactions of tunable strength. We analyze and model the contrast degradation and phase accumulation of the Ramsey signal and identify scaling laws for varying interrogation times, ensemble densities, and ensemble dimensionalities.
\end{abstract}


\maketitle

\section{Introduction}

\indent Strongly interacting many-body systems are at the focus of modern quantum physics~\cite{Bloch2008,Weimer2010}. Recent progress has seen the developing of techniques leading to the achievement of an outstanding level of control over quantum states and even nonequilibrium dynamics in systems exhibiting long-range interactions~\cite{Martin2013,Bettelli2013,Monroe2014,Jurcevic2014,Yan2013,Schauss2012,Schauss2015,Zeiher2016}. One of the extremely successful methods of investigating the effect of interactions in a many-body system is the Ramsey technique~\cite{Martin2013,Yan2013,Zeiher2016,Ramsey1950,Nipper2012} as it allows the monitoring of the evolution of coherences and correlations. This consists in an initialization process that brings the many-body system into a well-defined nonequilibrium state followed after some interrogation time by a mapping of the evolved many-body coherence into a population signal. The scanning of the interrogation (delay) time produces fringes with amplitude and phase strongly dependent on the characteristic interactions of the system.

\indent In this paper, we restrict our treatment to Ramsey interferometry applied to Rydberg atom ensembles~\cite{Alber1991,Noordam1992, Ohmori2009, Kozak2013, Zhou2014, Takei2015}. This occurs without loss of generality as we employ an Ising-type Hamiltonian applicable to a variety of distinct systems where only the particularities of the interaction strength differ~\cite{Worm2013, Feig2013, Hazzard2013, Hazzard2014, Mukherjee2015}.
The technique consists in the application of a first (pump) excitation pulse followed by a second (probe) pulse. For the particular example of the experiment described in Ref.~\cite{ Takei2015}, the pulses are obtained from a Michelson interferometer and are aimed at performing time-domain interferometry. An adjustable length of one interferometer arm allows one to set the delay $\tau$ between the pulses. This results in two identical pulses with envelopes $E(t)$ and $E(t-\tau)$. By scanning the delay between the pump and probe excitations we obtain the interferogram with fringes showing modulation at frequencies close to \textit{optical frequencies} (at the frequency difference between ground and Rydberg-excited levels). This is in contrast with standard Ramsey interferometry where the fringes occur with periodicity given by the difference between the laser and atomic frequencies. The signal is obtained by monitoring the population in the eigenstates, which in the absence of the interaction would oscillate at exactly the frequency difference between the initial and the final state. Most importantly, this technique allows the access of time scales in the picosecond to attosecond range, where the ultrafast electron dynamics can be probed directly~\cite{Yeazell1988,tenWolde1988,Yeazell1989,Takei2015}. By choosing sufficiently short pulses exhibiting a broad frequency spectrum, interactions at short internuclear distances are accessible and the blockaded regime ~\cite{Gould2004,Urban2009,Gaetan2009,Ates2007,Amthor2013,Takei2015} can be avoided.

\indent
As the key result of this manuscript we show that analytical results are possible for the general solution of the Ramsey signal of the interacting Rydberg system described by a many particle Ising model~\cite{Worm2013, Feig2013}. These results directly apply to experimental results~\cite{Takei2015}. This is done in a purely coherent regime where the delay $\tau$ between the pump and probe excitations is a few orders of magnitude shorter than the radiative decay or the mean free time between collisions in the ensemble. In particular, we show that the behavior of both contrast decay and phase shift of the signal can be derived essentially analytically under a continuous limit assumption. Our findings show that the decay of the coherence due to phase degradation depends strongly on the type of interaction and the dimension of the atomic distribution (we particularize to 1D, 2D, and 3D situations). The corresponding decay constants show particular density dependencies for different dimensions and types of interaction that can be directly tested in experiments.\\
\indent We proceed by introducing the details of the interactions in Rydberg ensembles in Sec.~\ref{Model} and the specifics of the Ramsey technique in Sec.~\ref{Ramsey interferometry in the time domain}. We then summarize general analytical results for the population signal in Sec.~\ref{Ramsey interferometry in the time domain} and compare them with mean-field results obtained in Sec.~\ref{Mean1}. The general formulas are then particularized for discrete and continuous configurations in Sec.~\ref{Lattice} and Sec.~\ref{Approx}, respectively.
\section{Model}
\label{Model}
Let us consider a simplified model for Rydberg atoms as two-level systems where the ground $|g\rangle$ and excited Rydberg state $|e\rangle$ are separated by an energy $\omega$ (in the following we set $\hbar$ to unity). For a generic atom indexed with $j$ we define projection operators for the excited state (spin-up)  $\hat{\sigma}_{e}^{(j)}=|e\rangle_{j}\langle e|_{j} = (1+\hat{\sigma}_{z}^{(j)})/2$, and the ground state (spin-down) $\hat{\sigma}_{g}^{(j)}=|g\rangle_{j}\langle g|_{j} = (1-\hat{\sigma}_{z}^{(j)})/2$. Transitions between levels are governed by lowering and raising operators defined as $\hat{\sigma}_{-}^{(j)}=|g\rangle_{j}\langle e|_{j}$ and its Hermitian conjugate $\hat{\sigma}_{+}^{(j)}=|e\rangle_{j}\langle g|_{j}$, respectively.
A freely evolving atom $j$ is then subjected to a Hamiltonian $H_{0} = \omega \hat{\sigma}_{z}^{(j)}/2$, with $\hat{\sigma}_{z}^{(j)} = \hat{\sigma}_{e}^{(j)} - \hat{\sigma}_{g}^{(j)}$.

Motivated by experiments~\cite{Takei2015}, we consider the following Hamiltonian for $N$ Rydberg atoms coupled by long-range ("all-to-all") interactions
\begin{eqnarray}
\label{Eq.2}
H = \sum_{j} \frac{\omega}{2} \hat{\sigma}_{z}^{(j)}+ \sum_{j,k} \frac{U_{jk}}{2}\hat{\sigma}_{e}^{(j)}\otimes \hat{\sigma}_{e}^{(k)},
\end{eqnarray}
where the interaction between atoms $j$ and $k$ is included in the term $U_{jk}$ (non-zero only for $k\neq j$). For the case of van der Waals interactions, $U_{jk}$ has the form $U_{jk} = C_{6}/{r_{jk}^{6}}$ where $r_{jk}$ is the separation between the atoms. In the following we will mainly consider this particular dependence of the interaction strength on distance. For illustration purposes we will focus on typical particle-particle separations of the order of $\mu$m and an interaction parameter $C_{6} = 2\pi\times 13.7\,$GHz$\times \mu$m$^{6}$, which has been measured for the 53D$_{3/2}$ Rydberg state of $^{87}$Rb atoms~\cite{Beguin2013}. In addition, in the continuous limit, where analytical results are possible, we will also consider the case of an effective dipole-dipole interaction given by $U_{jk} = C_{3}/{r_{jk}^{3}}$, which in a simplified approach governs the interaction between particles at close internuclear distances. Here, the measured parameter $C_{3} = 2\pi \times 2.1\,$GHz$\times \mu$m$^{3}$ for the 59D$_{3/2}$ Rydberg state of $^{87}$Rb has been taken into account for illustration purposes \cite{Ravets2014}. Finally, we will also consider a hybrid model where the interactions have a dipole-dipole character at distances below $r_{1}$ (a given crossover distance), while for distances larger than $r_{1}$ interactions are of the van der Waals type.

\section{General expression for the time-domain Ramsey signal}
\label{Ramsey interferometry in the time domain}

We start by illustrating the Ramsey procedure for a single atom. In a first step, the atom is driven into a superposition of ground and excited states with coefficients $c_{g}$ and $c_{e}$. This is achieved by a unitary $2 \times 2$ matrix transformation (denoted by $A$) applied to the initial state which is given by the ground state (spin down) $|g\rangle$, here identified with the vector $(1, 0)^{\top}$. The application of the matrix $A$ represents the application of the first (pump) pulse in the experiment. Up to a global phase the most general form for this matrix is $A_{11}=c_{g}$ and $A_{12}=ic_{e}$, $A_{22}=c^{*}_{g}$ and $A_{21}=ic^{*}_{e}$ (see Appendix A). The system is allowed to freely evolve at the natural frequency $\omega$ for a time $\tau$ after which driving described by transformation $A$ is applied again (corresponding to the probe pulse in the experiment), resulting in the final-state vector
\begin{eqnarray}
\label{eq:Ram1}
|\Psi(\tau)\rangle = A e^{-iH_{0}\tau}A|g\rangle.
\end{eqnarray}
In experiments, the observable is given by the Ramsey signal that quantifies the Rydberg-state population $P(\tau) = \langle \Psi(\tau)|e\rangle \langle e|\Psi(\tau)\rangle$. The latter has in general a sinusoidal profile as a function of the delay time $\tau$.
Defining $p_{g} = c^{*}_{g}c_{g}$ and $p_{e} = c^{*}_{e}c_{e}$, the signal for a single atom can be readily expressed as  (see Appendix A)
\begin{eqnarray}
\label{eq:Ram2}
P(\tau) = 2p_{g}p_{e}\left[1 + \cos(\omega\tau+ \phi) \right],
\end{eqnarray}
where $\phi$ (defined as the phase of the complex $c_g$) is a trivial ac-Stark shift acquired during the pulse excitation.\\
\indent For an ensemble of interacting Rydberg atoms, we make the assumption that the duration of the Ramsey pulses is short compared to the characteristic interaction time defined as the inverse of the particle-particle interaction strength at the mean inter particle distance. This is typically fulfilled in experiments with field-induced interactions~\cite{Nipper2012,Zeiher2016} or ultrafast lasers~\cite{Takei2015}. In such a case we can neglect the effect of interactions during the pulse and can immediately generalize Eq.~\eqref{eq:Ram1} to the many atom case by writing the state of the system right before the application of the probe pulse as
\begin{eqnarray}
|\Phi(\tau)\rangle_{N} = e^{-iH \tau}A^{\otimes N}|g\rangle^{\otimes N}.
\end{eqnarray}
By calculating the corresponding density operator $\hat{\rho}_{1\dots N} = |\Phi(\tau)\rangle_{N}\langle \Phi(\tau)|_{N}$ and taking traces over $N-1$ particles except particle $j$ we obtain the following analytical expression for the {\it single-particle} density operator (see Appendix B):
\begin{eqnarray}
\label{AEq.10}
\hat{\rho}_{j} &=& p_g\hat{\sigma}_{g}^{(j)} - i\prod_{k \neq j}^{N}\left(p_{g} + e^{iU_{jk}\tau}p_{e}\right) e^{i\omega\tau}c_{g}c_{e}\hat{\sigma}_{-}^{(j)} \\\nonumber
& & + i\prod_{k \neq j}^{N}\left(p_{g} + e^{-iU_{jk}\tau}p_{e}\right) e^{-i\omega\tau}c^{*}_{e}c^{*}_{g}\hat{\sigma}_{+}^{(j)} + p_e\hat{\sigma}_{e}^{(j)}.
\end{eqnarray}

The population in the excited state $|e\rangle$ following the application of the probe pulse is then described by $P_{j}(\tau) = \langle e|A\hat{\rho}_{j}A^{*}|e \rangle$ for a single atom. As shown in Appendix B we then obtain the signal:
\begin{equation}
\label{c}
P_{j}(\tau) = 2p_{g}p_{e}\Re \left\{ 1+ \alpha(\tau) G_j(\tau) \right\},
\end{equation}
with $\alpha(\tau)=e^{i(\omega \tau+\phi)}$ and the information acquired from the interactions is contained in the following term:
\begin{equation}
\label{AEq.11}
G_j(\tau) = \prod_{k \neq j}\left( p_{g} + p_{e}e^{iU_{jk}\tau}\right).
\end{equation}
Notice that $G_j(\tau)$ is an interaction-induced modulation of the signal that directly reflects the coherences established in the system right before the probe pulse is applied [see the term $\prod_{k \neq j}^{N}\left(p_{g} + e^{iU_{jk}\tau}p_{e}\right) e^{i\omega\tau}c_{g}c_{e}$ in Eq.~\eqref{AEq.10}]. This kind of expression has been previously presented in the framework of a general Ising model in~\cite{Worm2013, Feig2013, Hazzard2013, Hazzard2014} and is fully derived in Appendix B. The signature of the interactions is fully contained in $G_j(\tau)$, and in the following sections we will monitor the contrast degradation and phase accumulation of the Ramsey signal defined as:
\begin{eqnarray}
\label{eq:4}
\mu_j(\tau)&=&| G_j(\tau)| \;\;\;\;\;\;\;\;\;\;\;\;\;\;\;\;\;\;\;\;\;\;  \mathrm{ contrast}\\
\nu_j(\tau)&=& -i \ln \left(G_j(\tau)/|G_j(\tau)|\right) \;\;\;\mathrm{ phase}.
\end{eqnarray}
Notice that in the absence of interactions (i.e., for independent particles) the function $G_j(\tau)$ converges to unity and the population has the same expression as in the single-particle case [Eq.~\eqref{eq:Ram2}] predicting no contrast degradation ($\mu_j(\tau)=1$) or phase accumulation ($\nu_j(\tau)=0$). These two quantities are central as they can be experimentally accessed~\cite{Takei2015} and we will evaluate them both for discrete and continuous distributions.\\
\indent We notice that so far we have treated the case of the signal obtained for a given particle $j$ (i.e., \textit{single particle level}). 
The experimentally observed quantity is of course a \textit{many particle} signal obtained from the averaging over contributions coming from all individual atoms:
\begin{equation}
\label{pop}
P(\tau) = \frac{1}{N}\sum^{N}_{j=1}P_{j}(\tau).
\end{equation}
This latter averaging will be important in the case of inhomogeneous distributions of Rydberg atoms, such as the case of finite atomic samples in the presence of external magnetic or optical quadratic confining potentials.

\section{Comparison to the linearized approach (mean-field model)}
\label{Mean1}

\begin{figure}[t]
	\begin{center}
		\includegraphics[width=0.45\textwidth]{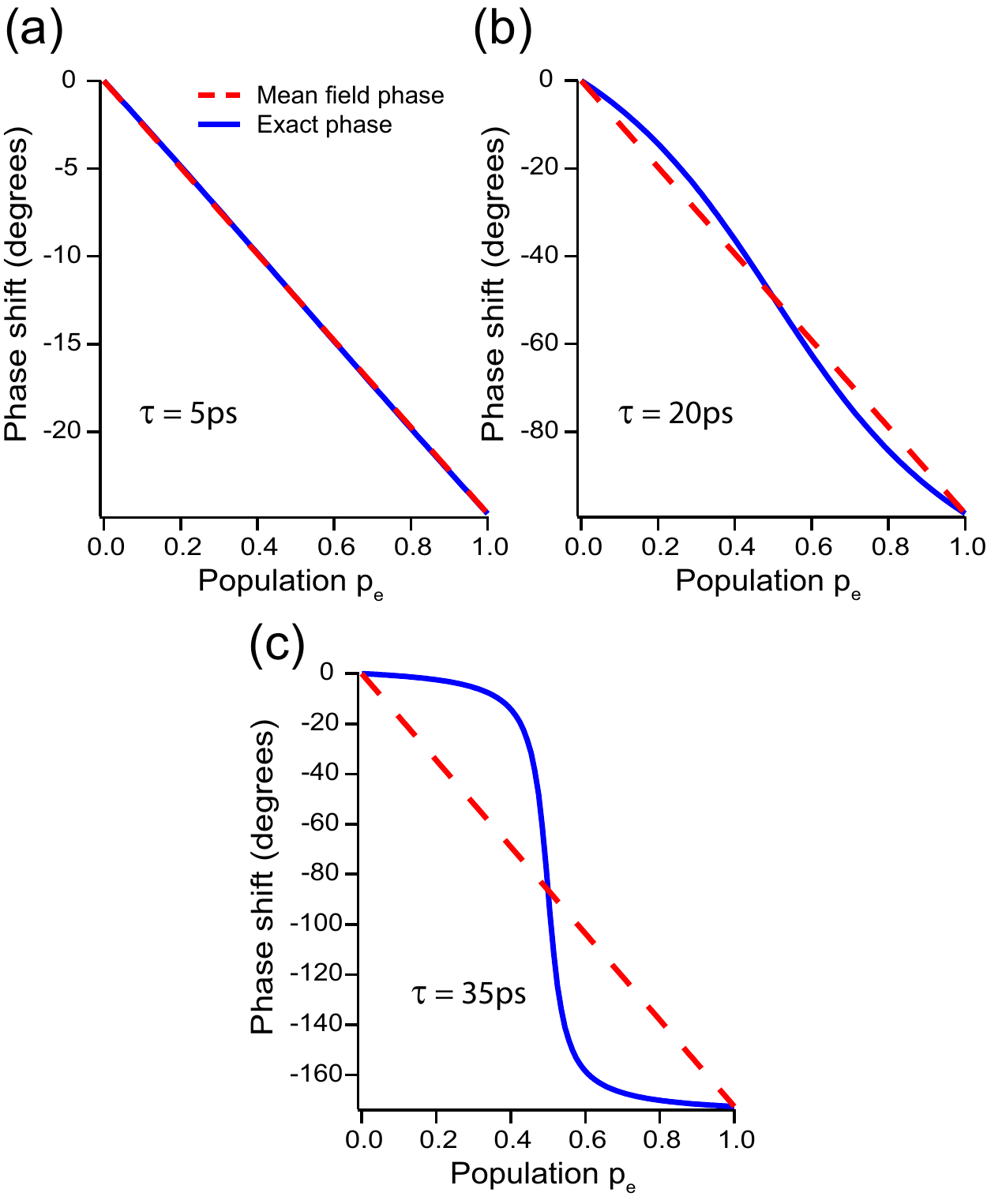}
		\caption{{\bf Comparison between the mean-field and exact-model phase shift.}
In (\textbf{a}), (\textbf{b}) and (\textbf{c}) the phase shift obtained from the mean-field solution (dashed line) and the exact model solution (solid line) for two interacting atoms at a distance of $1\,\mu$m is plotted against the population in the excited state $p_{e}$ and  presented for a pump-probe delay of $\tau = 5\,$ps, $\tau = 20\,$ps and $\tau = 35\,$ps, respectively. The interaction strength is given by $C_{6} = 2\pi\times 13.7\,$GHz$\times\mu$m$^{6}$ \cite{Beguin2013}. Only at early delay times $\tau$ both solutions agree. Irrespective of the population $p_{e}$ the two solutions deviate strongly from each other with increasing delay time.}
		\label{fig.0}
	\end{center}
\end{figure}
We now introduce a mean-field approach and compare its validity to the full many-body solution obtained in the previous section. We first shift the full Hamiltonian of Eq.~\eqref{Eq.2} by a constant energy term and proceed by the usual assumption characteristic of mean-field techniques that the nonlinear terms can be linearized around some common value $\langle \hat{\sigma}_{e} \rangle $ by ignoring smaller contributions and replace $\hat{\sigma}_{e}^{(j)} \hat{\sigma}_{e}^{(k)}$ with $\langle \hat{\sigma}_{e} \rangle (\hat{\sigma}_{e}^{(j)}+ \hat{\sigma}_{e}^{(k)}-\langle \hat{\sigma}_{e} \rangle )$. The linearized Hamiltonian in a separable form becomes $H^{\mathrm{mf}} = \sum_{j} H_{j}^{\mathrm{mf}}$, where
\begin{eqnarray}
\label{AEq.15}
H_{j}^{\mathrm{mf}} = \left[ \omega + \left( 1 - \frac{\langle \hat{\sigma}_{e} \rangle}{2}  \right)\Delta\omega_{j} \right]  \hat{\sigma}_{e}^{(j)} - \frac{\langle \hat{\sigma}_{e} \rangle \Delta\omega_{j}}{2} \hat{\sigma}_{g}^{(j)}, \;\;\;\;\;\;
\end{eqnarray}
with $\Delta\omega_{j} = \sum_{k} U_{jk}$ defined as a frequency shift of atom $j$ owed to the background field produced by all the other atoms. We now analyze the dynamics of particle $j$ where $\langle \hat{\sigma}_{e} \rangle=p_{e}$ to obtain
\begin{equation}
\label{AEq19}
G^{\mathrm{mf}}_j(\tau) = \prod_{k \neq j}e^{i p_{e} U_{jk}\tau}.
\end{equation}
\noindent \textbf{Contrast degradation} As obvious from the above expression that predicts a unit modulus for $G^{\mathrm{mf}}_j(\tau)$ at all times, the mean-field approach predicts no contrast degradation for individual particles. This is in contrast to the full solution prediction which allows degradation even at the \textit{single-particle level}. This becomes evident by recasting Eq.~\eqref{AEq.11} in the following form:
\begin{eqnarray}
\label{AEq.20}
\nonumber
G_j(\tau) &=&  \prod_{k \neq j}\sqrt{1-4p_{g} p_{e}\sin^{2}\left( \frac{U_{jk}\tau}{2}\right)} \\
&\times &   e^{i\sum_{k} \frac{U_{jk}\tau}{2} + \arctan\left((p_{e}-p_{g})\tan\left( \frac{U_{jk}\tau}{2}\right) \right)}.
\end{eqnarray}
Each term in the product has a modulus less than $1$ thus leading to a total amplitude reduction of the $G_j(\tau)$ function at nearly all times.

Generalizing to the \textit{many-particle level} we see that the mean-field solution can also not account for contrast degradation in the case of regularly spaced ensembles (e.g., atoms trapped in an optical lattice with even interatomic spacings). This follows from the fact that the ensemble-averaged signal includes the same $\Delta\omega_{j}\tau$ for all atoms $j=1,...N$ and the many particle signal becomes effectively a single particle signal. However, disordered ensembles can show degradation in the mean field approach after performing an averaging over all atoms; this comes from the fact that for a large degree of randomness in the inter-particle spacing $G^{\mathrm{mf}}_j(\tau) \neq G^{\mathrm{mf}}_k(\tau)$ for $j \neq k$ can lead to destructive interference in the sum over $G^{\mathrm{mf}}_j(\tau)$.\\

\begin{figure*}[ht]
	\begin{center}
		\includegraphics[width=0.70\textwidth]{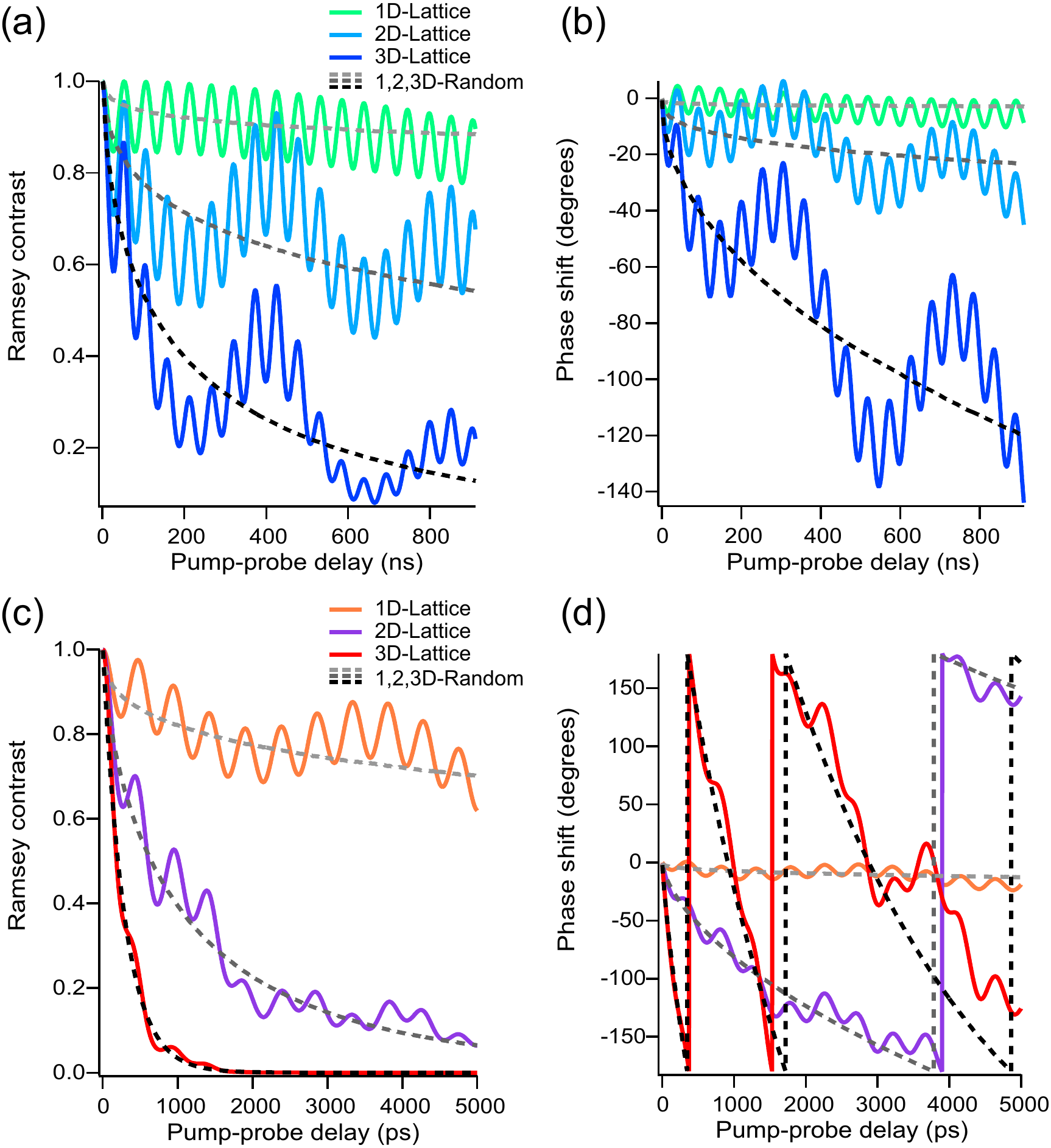}
		\caption{{\bf Ramsey interferogram for a one-, two- and three-dimensional lattice.}
In (\textbf{a}) and (\textbf{c}) the Ramsey contrast is shown for van der Waals  and dipole-dipole interacting atoms where each atom has a population of $p_{e} = 4\,\%$ in the Rydberg state. The Ramsey contrast is compared to the Ramsey contrast of a randomized homogeneous atom distribution of the same dimension, density and population, which has been obtained from Eq.~\eqref{AEq.11} using a Monte Carlo simulation for the atomic distribution. The lattice constant is $a = 3\,\mu$m for the van der Waals and $a = 1\,\mu$m for the dipole-dipole interaction scenario. The interaction strengths are defined by $C_{6} = 2\pi \times 13.7\,$GHz$\times \mu$m$^{6}$ (a,b) and $C_{3} = 2\pi \times 2.1\,$GHz$\times \mu$m$^{3}$ (c,d), which are obtained from \cite{Beguin2013} for the 53D$_{3/2}$ Rydberg level and \cite{Ravets2014} for the 59D$_{3/2}$ Rydberg level, respectively. The number of atoms considered for each simulation is given by $N = 729 = 27 \times 27 = 9 \times 9 \times 9$. In (\textbf{b}) and (\textbf{d}) the corresponding phase signals for the van der Waals and dipole-dipole interaction are presented, respectively. }
		\label{fig.1}
	\end{center}
\end{figure*}

\noindent \textbf{Phase accumulation} We now focus on the evolution of the accumulated phase for different time delays and excited-state populations. The comparison between mean-field (dashed line) and the full solution (continuous line) predictions is illustrated in Fig.~\ref{fig.0} for three different times. At times smaller than the inverse of the characteristic interaction time scale [$\tau < \min_{k} (1/U_{jk})$] the two solutions coincide in the whole population range. This is easily explained via a linearization of the expressions in the argument of Eq.~\eqref{AEq.20}
\begin{eqnarray}
\label{AEq.21}
G_j(\tau) &\approx&  G^{\mathrm{mf}}_j(\tau) \left(1-\frac{p_{g}p_{e}}{2}\prod_{k \neq j}\left( U_{jk}\tau\right)^{2}\right),
\end{eqnarray}
showing perfect agreement for the predicted phase accumulation. The disagreement in predicted phases occurs with increasing $\tau$, for essentially all $\tau$.

\section{Discrete distributions}
\label{Lattice}

Let us now consider ordered atomic configurations in $d$ dimensions ($d=1,2,3$), which can be realized for example by trapping Rydberg atoms by optical means~\cite{Saffman2016, Schauss2012,Zeiher2015, Saffman2013, Browaeys2016} or in arrays of magnetic traps~\cite{Whitlock2016a, Whitlock2016b}. For lattice configurations of lattice constant $a$ we evaluate the results of the theory developed in the previous section particularized to isotropic van der Waals and dipole-dipole interactions.\\
\indent We assume that the sample is large and each atom is equivalent to its neighbors. This allows us to reduce the whole problem to that of a single given atom surrounded by $N-1$ atoms. In a first step we numerically evaluate the expression in Eq.~\eqref{AEq.11} for three lattices containing the same number of atoms $N=729$ in one, two ($27 \times 27 $), and three ($9 \times 9 \times 9$) dimensions. We then plot the Ramsey contrast degradation and phase accumulation signals in Fig.~\ref{fig.1}. The contrast and phase signals show characteristic oscillations. By inspection, we find that the latter originate essentially from the interaction with the nearest and next nearest neighbors. The trend of the oscillatory curves agrees very well with the results obtained from a numerical simulation of Eq.~\eqref{AEq.11} averaged over a randomized disordered system of the same density and atom number. In this latter case, however, averaging results in a loss of the fast oscillations.\\
\indent In the 1D situation, owing to the rapid fall-off of the interaction strength with distance in the van der Waals interaction case, a nearest-neighbor treatment seems to be sufficient to describe the dynamics (as inferred from the single frequency oscillation of the upper green curves in Fig.~\ref{fig.1}a and Fig.~\ref{fig.1}b). This is not given in the dipole-dipole interaction scenario displayed in Fig.~\ref{fig.1}c and Fig.~\ref{fig.1}d, where contributions from the next-nearest neighbors enter significantly earlier due to the less rapid fall-off of the interaction strength with distance for the dipole-dipole interaction in comparison with the van der Waals interaction case. Especially for higher dimensions, the plotted behavior shows beatings of competing frequencies, clearly suggesting that a simplified model including both the nearest and the next-nearest neighbors does not suffice. Similar oscillations have been presented and discussed in~\cite{Yan2013, Hazzard2014a}.

\section{Continuous distributions}
\label{Approx}

Let us now consider a general formulation for disordered distributions of atoms that allows us to perform a transition to the continuum to find elegant expressions for Eq.~\eqref{AEq.11} particularized for 1D, 2D, and 3D situations. This formulation is geared towards the treatment of ultracold gases in dipole traps~\cite{Grimm2000,Takei2015}; however, it can also be applied to regular arrangements of atoms with a controlled level of disorder (such as for atoms trapped in optical lattices). As before, we consider a simplified but realistic scenario (applicable, for example, to experiments with ultrafast lasers~\cite{Takei2015}) where the delay $\tau$ between the pump and probe excitations is a few orders of magnitude shorter than the radiative decay or the mean free time between collisions in the ensemble. This allows us to focus on the Ramsey signal for the coherent part of the dynamics of the interacting spins, for which we derive analytic expressions.\\

The key point in our procedure is the expansion of the product in Eq.~\eqref{AEq.11} into sums, followed by the transformation of sums into integrals allowed by the continuous distribution limit. In the following, for simplicity we will refer to the particle of interest by the index $j=1$ (clearly the results are independent of this labeling). The binomial expansion reads
\begin{widetext}
\begin{eqnarray}
\label{eq:Approx2}
G_1(\tau) &=& p_{g}^{N-1} + p_{g}^{N-2}p_{e} \left( \sum_{i_{1}=2}^{N}e^{iU_{1i_{1}}\tau} \right) + p_{g}^{N-3}p_{e}^{2} \left( \sum_{i_{1}=2}^{N-1} \sum_{i_{2} > i_{1}}^{N} e^{i(U_{1i_{1}} +  U_{1i_{2}})\tau} \right) \\\nonumber
&+& \cdots \\\nonumber
&+&  p_{e}^{N-1} \left( \sum_{i_{1}=2}^{N-(N-2)} \cdots \sum_{i_{(N-1)} > i_{(N-2)}}^{N} e^{i(U_{1i_{1}} + \dots +  U_{1i_{(N-1)}} )\tau} \right).
\end{eqnarray}
\end{widetext}
\noindent We have assumed that the atoms are arranged as a function of increasing separation $|\textbf{r}_j|$ from the atom of interest (atom $1$ situated in the origin) and can be indexed by a number $j$ running from $2$ to $N$. A generic term in the expression Eq.~\eqref{eq:Approx2} containing contributions from $m$ neighboring atoms will contain $m$ sums running over a set of $m$ indexes denoted by $i_1,\cdots i_m$.

We now perform a transition to the continuum where the sums in Eq.~\eqref{eq:Approx2} are substituted by integrals: $\sum_{i_1,i_2\cdots i_m}\cdots\rightarrow \int_{V_{i_1}} \cdots \int_{V_{i_m}} dV_{i_1} \cdots dV_{i_m} f_{m}\cdots$. The term of rank $m$ is properly weighed with a conditional distribution function $f_{m}$ which depends on the characteristic of the system: for example we will contrast homogeneous random distributions (such as those usually found with ultracold atomic gases) to regular distributions (such as those resulting from atoms trapped in an optical lattice). The normalization condition with a general usefulness (both for random as well as ordered distributions) is immediately obtained by substituting unity in the above sums-integrals transformation to obtain:
\begin{eqnarray}
\label{eq:Approx4}
\int_{V_{1}} \cdots \int_{V_{m}} dV_{1} \cdots dV_{m} f_{m}  = \binom{N-1}{m},
\end{eqnarray}
and making the observation that the implicit ordering of the indexes in the sum of rank $m$ leads to $\sum_{i_1,i_2...i_m} 1=\binom{N-1}{m}$.
Particularizing for example to regular arrays with small levels of fluctuations in the equilibrium positions, the distribution functions that fulfill the normalization condition can be approximated with Gaussians centered around the atomic position with a waist reflecting the positioning uncertainty $f_m=\sum_{i_1,i_2...i_m} \prod_{i_k} (\sqrt{\pi}w)^{-d} e^{-(\textbf{r}-{\textbf{r}_{i_k}})^2/w^2}$. In the purely ordered case, one can take the limit $w\rightarrow 0$ which leads to $f_m=\sum_{i_1,i_2...i_m} \prod_{i_k} \delta(\textbf{r}-\textbf{r}_{i_k})$. This reproduces the situation of a 'frozen' optical lattice and the continuous and discrete descriptions coincide.\\
\indent As a next particular case which we will focus on in the following sections, we will consider the homogeneous density distributions within a volume $V_0=N_0 n(\textbf{R})$ around the atom indexed by $j=1$ with position now shifted from the origin to $\textbf{R}$. Here $N_0$ is the number of atoms in the volume $V_0$ around atom $1$ that participate with a non-negligible contribution to $G_1(\tau)$. In experiments, this cut-off volume $V_0$ is a function of the total interrogation time: for longer $\tau$ more and more atoms contribute a considerable shift $U(r,\theta,\phi)\tau$
to the Ramsey signals. 
In this case, for an ensemble in $d$ dimensions we can approximate the function $f_{m}$ locally with a constant value $f_{m} \approx 1/V_{0}^m \binom{N_{0}-1}{m}$. This description is valid for terms with $m \ll N_{0}$. The homogeneity condition also imposes the restriction on density variations $ |\nabla n(\textbf{R})| r_0/n(\textbf{R})\ll 1$, where $r_{0}$ is the radius determined from the local spherical volume $N_{0}/n=V_0 = \pi^{d/2}r_{0}^{d}/\Gamma(d/2+1)$.\\

Let us now further simplify our treatment to consider an isotropic case, such that the integrals over angular dependence are trivial. We furthermore introduce a (short-range) Rydberg-blockade volume with radius $r_\mathrm{B} \ll r_0$ that determines a short-range limit in the integration range, which is physically motivated by the finite (however large) bandwidth of the driving laser in experiments~\cite{Takei2015}. Under these conditions, the first sum of the expansion in Eq.~\eqref{eq:Approx2} turns into a fairly simple integral:
\begin{eqnarray}
\label{int}
\gamma(\tau)=\frac{d}{r_{0}^{d}-r_{\mathrm{B}}^{d}} \int_{r_{\mathrm{B}}}^{r_{0}}dr r^{d-1} e^{iU(r)\tau},
\end{eqnarray}
where we have restricted the discussion here and in the following to the treatment of isotropic interaction potentials. A general treatment for anisotropic van der Waals and dipole-dipole interactions is presented in Appendix D, which shows that the solutions for $\gamma(\tau)$ agree up to a prefactor with the ones derived for isotropic potentials when $r_{\mathrm{B}} = 0$. 
The next terms are in general harder to evaluate. However, in the limit of large $N_{0}$, we can ignore the different integration boundaries and evaluate a term of rank $m$ as $\binom{N_{0}-1}{m}\gamma(\tau)^m$. We then observe that Eq.~\eqref{eq:Approx2} can be recast in a binomial form:
\begin{eqnarray}
\label{eq:Approx5}
\label{Gcont}
G(\tau) &\approx& \left( p_{g} + p_{e}\gamma(\tau)\right)^{N_{0}-1}.
\end{eqnarray}

\begin{figure}[t]
	\begin{center}
		\includegraphics[width=0.49\textwidth]{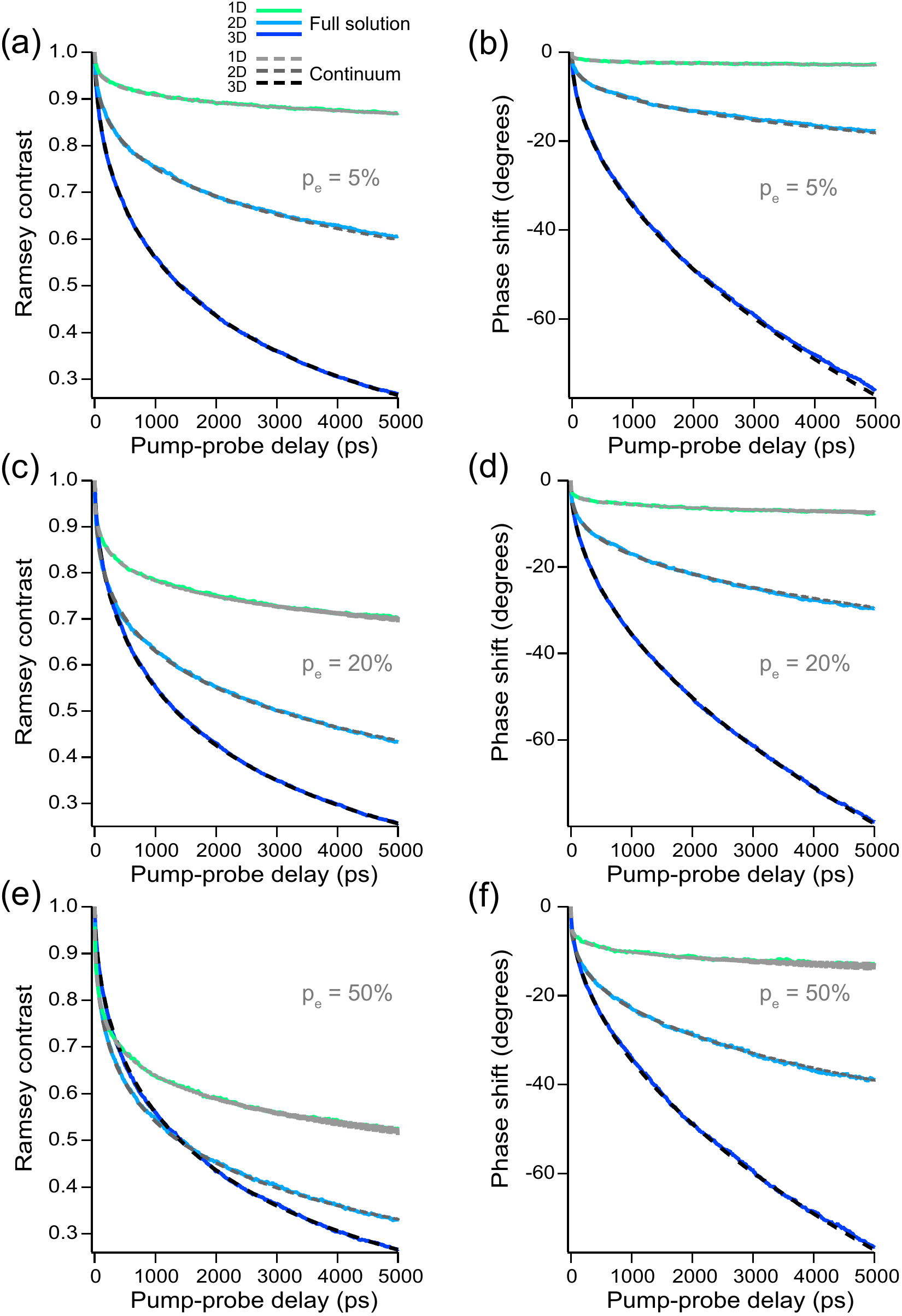}
		\caption{{\bf Ramsey interferogram for a homogeneous atom distribution.} In (\textbf{a}), (\textbf{c}) and (\textbf{e}) the Ramsey contrast and in (\textbf{b}), (\textbf{d}) and (\textbf{f}) the Ramsey phase is presented for $p_{e} = 5\,\%$, (a,b), $p_{e} = 20\,\%$, (c,d) and $p_{e} = 50\,\%$ (e,f). The interaction strength is given by $C_{6} = 2\pi\times 13.7\,$GHz$\times\mu$m$^{6}$ \cite{Beguin2013} and densities are adjusted for comparison. The exact solutions from Eq.~\eqref{AEq.11} are given by the solid lines for $N=5650$ atoms, while the gray dashed lines following Eq.~\eqref{eq:9da} (contrast) and Eq.~\eqref{eq:9db} (phase) show the solutions derived by the continuum approximation for $N \rightarrow \infty$.}
		\label{fig.2}
	\end{center}
\end{figure}

This is a main result of our analysis as it allows us on one hand to perform faster numerical simulations and on the other hand to extract scaling laws for the contrast degradation and phase accumulation with density and interrogation time. We proceed by analyzing different interaction regimes characterized by $\sim r^{-6}$ (van der Waals) or by $\sim r^{-3}$ (dipole-dipole) scalings as well as a hybrid regime.

\subsection{Van der Waals interactions}

\indent For the van der Waals interaction regime described by $U(r) = C_{6}/r^{6}$ we can simplify the function $\gamma(\tau)$, by evaluating the integral in Eq.~\eqref{int} in the limit of $r_\mathrm{B}$ approaching zero. Using a set of transformations and the asymptotes of the general Fresnel integrals~\cite{Mathar2012} (see details in Appendix C) we find a general form holding for any dimension $d \in \{1,2,3\}$:
\begin{eqnarray}
\label{eq:9d}
\gamma^{d\mathrm{D}}_{\mathrm{vdW}}(\tau) &\approx& 1 + \frac{i \sigma_{d}\Gamma\left(\frac{6-d}{6}\right)}{N_{0}}e^{\frac{i(6-d)\pi}{12}}\tau^{d/6},
\end{eqnarray}

\indent After the replacement of the above expression into Eq.~\eqref{Gcont} and using $p_g +p_e=1$, we notice that in the limit $N_{0}\rightarrow\infty$ the expression coincides with an exponential from which we can now immediately deduce the corresponding decay of the contrast $\mu(\tau)$ in an arbitrary dimension $d$ as
\begin{eqnarray}
\label{mu}
\mu^{d\mathrm{D}}_{\mathrm{vdW}}(\tau) &\approx& e^{ -p_{e}\sigma_{d}\Gamma\left(\frac{6-d}{6}\right) \sin\left(\frac{(6-d)\pi}{12}\right)\tau^{d/6}}.
\end{eqnarray}

\noindent The phase of the Ramsey signal is given by
\begin{eqnarray}
\label{eq:9d}
\nu^{d\mathrm{D}}_{\mathrm{vdW}}(\tau) &\approx& p_{e}\sigma_{d}\Gamma\left(\frac{6-d}{6}\right) \cos\left(\frac{(6-d)\pi}{12}\right)\tau^{d/6}. \;\;
\end{eqnarray}
From Eq.~\eqref{mu} we observe that the exponential decay follows with the sixth root $e^{-\tilde{\alpha}_{1}\sqrt[6]{\tau}}$, the cube root $e^{-\tilde{\alpha}_{2}\sqrt[3]{\tau}}$ and the square root $e^{-\tilde{\alpha}_{3}\sqrt{\tau}}$ of the delay time $\tau$ for one, two and three dimensions, respectively.
Also, for the decay time $\tau_{\mathrm{dec}}$, which is defined by $\mu^{d\mathrm{D}}_{\mathrm{vdW}} = 1/e$ we find the relation
\begin{eqnarray}
\label{eq:9de}
\tau_{\mathrm{dec}} \propto \frac{1}{n^{6/d}},
\end{eqnarray} which displays the characteristic density dependence of the decay for different dimensions for van der Waals interacting atoms.

Including a finite blockade radius (and corresponding blockade energy $\omega_{\mathrm{B}}=C_6/r^6_{\mathrm{B}}$) allows us to deduce more general expressions for the contrast and phase function as
\begin{eqnarray}
\label{eq:9da}
\nonumber
\mu^{d\mathrm{D}}_{\mathrm{vdW}}(\tau,\omega_{\mathrm{B}}) &\approx& e^{-p_{e}\sigma_{d}\left[\frac{\cos(\omega_{\mathrm{B}}\tau)-1}{\omega_{\mathrm{B}}^{d/6}}+\frac{6}{d}\tau^{d/6}\tilde{S}_{d}\left((\omega_{\mathrm{B}}\tau)^{d/6}\right)\right]}.\\
\end{eqnarray}
and
\begin{eqnarray}
\label{eq:9db}
\nonumber
\nu^{d\mathrm{D}}_{\mathrm{vdW}}(\tau,\omega_{\mathrm{B}}) &\approx& -p_{e}\sigma_{d}\frac{\sin(\omega_{\mathrm{B}}\tau)}{\omega_{\mathrm{B}}^{d/6}}+  \\
& & + p_{e}\sigma_{d}\frac{6}{d}\tau^{d/6}\tilde{C}_{d}\left((\omega_{\mathrm{B}}\tau)^{d/6}\right),
\end{eqnarray}
where we have used the following Fresnel integrals $\tilde{S}_{d}(x) =  S_{(3-d)^{2},6/d}(x) = \int_{0}^{x}t^{(3-d)^{2}}\sin(t^{6/d})dt$ and $\tilde{C}_{d}(x) = C_{(3-d)^{2},6/d}(x) = \int_{0}^{x}t^{(3-d)^{2}}\cos(t^{6/d})dt$. In Fig.~\ref{fig.2} we compare these results with those derived from a Monte Carlo simulation using the exact solution presented in Eq.~\eqref{AEq.11} for three different populations in the excited state. The densities are adjusted for comparison and are defined by the characteristic length $a$ with $n = 1/a^{d}$. Here, we have chosen $a = 1.6\,\mu$m, $a = 2.5\,\mu$m, and $a = 3.4\,\mu$m for $p_{e} = 5\,\%$, $p_{e} = 20\,\%$, and $p_{e} = 50\,\%$, respectively. The Monte Carlo simulation is constrained by the condition to form a homogenous density distribution in a sphere of size $N/n$ with $N = 5650$ atoms and $n$ being the constant density. By calculating the distance to the center of the sphere we obtain the interaction contribution $C_{6}/r^{6}$ from each particle. Additionally, we average over $1000$ simulations for $p_{e} = 5\,\%$, $5000$ simulations for $p_{e} = 20\,\%$, and $10000$ simulations for $p_{e} = 50\,\%$ to obtain the contrast and phase signals displayed in Fig.~\ref{fig.2}.

\subsection{Dipole-dipole interactions}

\indent Similar to the van der Waals interaction, we define now a blockade energy $\omega_{\mathrm{B}} = C_{3}/r^{3}_{\mathrm{B}}$. In 1D we obtain a simple result showing that $\gamma^{\mathrm{1D}}_{\mathrm{DD}}$ is equivalent to $\gamma^{\mathrm{2D}}_{\mathrm{vdW}}$, when $\sigma_{2}$ is exchanged with $\kappa_{1} = 2n\sqrt[3]{C_{3}}$, (see Appendix C). For higher dimensions we find the following (more involved) expressions:
\begin{eqnarray}
\label{eq:9x}
\nonumber
\gamma^{\mathrm{2D}}_{\mathrm{DD}} &\approx& 1 + \frac{\kappa_{2}}{\omega^{2/3}_{\mathrm{B}}N_{0}}\left[1 - \left(1-3i\omega_{\mathrm{B}}\tau)e^{i\omega_{B}\tau}\right)\right]+ \\
& & + \frac{9\kappa_{2}\tau^{2/3}}{N_{0}}\left[C_{3,3}(\sqrt[3]{\omega_{\mathrm{B}}\tau})+iS_{3,3}(\sqrt[3]{\omega_{\mathrm{B}}\tau}) \right], \;\;\;
\end{eqnarray}
where $\kappa_{2} = \pi n(C_{3})^{2/3}$ and notice that $\omega_{\mathrm{B}}$ needs to remain finite to avoid divergence. This results in

\begin{eqnarray}
\label{eq:9d}
\nonumber
\mu^{\mathrm{2D}}_{\mathrm{DD}}(\tau) &\approx& e^{-p_{e}\kappa_{2}\left[\frac{\cos(\omega_{\mathrm{B}}\tau)+3\omega_{\mathrm{B}}\tau\sin(\omega_{\mathrm{B}}\tau)-1}{\omega_{\mathrm{B}}^{2/3}}-9\tau^{\frac{2}{3}}C_{3,3}\left(\sqrt[3]{\omega_{\mathrm{B}}\tau}\right)\right]}\\
\end{eqnarray}
and
\begin{eqnarray}
\label{eq:9d}
\nonumber
\nu^{\mathrm{2D}}_{\mathrm{DD}}(\tau) &\approx& -\frac{p_{e}\kappa_{2}}{\omega_{\mathrm{B}}^{2/3}}\left[\sin(\omega_{\mathrm{B}}\tau)-3\omega_{\mathrm{B}}\tau\cos(\omega_{\mathrm{B}}\tau)\right]+\\
& &
 +9p_{e}\kappa_{2}\tau^{\frac{2}{3}}S_{3,3}(\sqrt[3]{\omega_{\mathrm{B}}\tau}),
\end{eqnarray}
showing a mixture of an exponential decay with a linear $e^{-\hat{\alpha}_{1}\tau}$ and a square of a cube root $e^{\hat{\alpha}_{2}\tau^{2/3}}$ component.

For the solution in three dimensions we use the integrals $\mathrm{Si}(x) = \int_{0}^{x}dt\sin(t)/t$ and $\mathrm{Ci}(x) = -\int_{x}^{\infty}dt\cos(t)/t$ and we obtain
\begin{eqnarray}
\label{eq:9z}
\gamma^{\mathrm{3D}}_{\mathrm{DD}} &\approx& 1 + \frac{\kappa_{3} \tau}{N_{0}} \left[i\left(1-\gamma - \ln \frac{\kappa_{3}\tau}{N_{0}}\right) -\pi/2\right]
\end{eqnarray}
for the dipole-dipole interaction with $\kappa_{3} = \frac{4\pi n C_{3}}{3}$, the Euler-Mascheroni constant $\gamma$, and $\omega_{\mathrm{B}} \rightarrow \infty$.
The function $G_1(\tau)$ for the dipole-dipole interaction is given by
\begin{eqnarray}
\label{eq:10}
\nonumber
G^{\mathrm{3D}}_{\mathrm{DD}}(\tau)
&\approx& \left\{ 1 + \frac{p_{e}\kappa_{3} \tau}{N_{0}} \left[i\left(1-\gamma - \ln \frac{\kappa_{3}\tau}{N_{0}}\right) -\pi/2\right]\right\}^{N_{0}}.\\
\end{eqnarray}
The term $\ln(\kappa_{3}\tau/N_{0})$ leads to a divergent oscillatory part for increasing $N_{0}$ in the expression enclosed by brackets. The decay amplitude, on the other hand, is well defined and remains finite. Ignoring the oscillatory part we obtain $[1 - \pi p_{e}\kappa_{3}\tau /(2 N_{0})]^{N_{0}}$, which results in the limit of large $N_{0}$ in
\begin{eqnarray}
\label{eq:10a}
\mu^{\mathrm{3D}}_{\mathrm{DD}}(\tau) &=& e^{-\frac{\pi}{2}p_{e}\kappa_{3}\tau}.
\end{eqnarray}
This shows that the coherence follows an exponential decay with linear dependence in the time delay and a decay constant proportional to the density and the interaction strength $C_{3}$.

\subsection{Hybrid interaction regime}

\indent With the solutions of the integrals in Eq.~\eqref{eq:2} and  Eq.~\eqref{eq:1aa} in Appendix C for the dipole-dipole and van der Waals potentials,  respectively, we can find a solution for a hybrid potential that has a spatial dependence proportional to $r^{-3}$ below a given radius $r_1$ and to $r^{-6}$ above $r_1$, with $\omega_{1} =  C_{3}/r_{1}^{3} = C_{6}/r_{1}^{6}$. This choice is an approximation of realistic interaction potentials with Rydberg atoms~\cite{Takei2015}. Here, this is exemplified for three dimensions.
\noindent For $\omega_{\mathrm{B}} \rightarrow \infty$ the resulting expression is
\begin{eqnarray}
\label{eq:19}
\nonumber
\gamma^{\mathrm{3D}}_{\mathrm{Hyb}}(\tau) &=& 1 - i\frac{\kappa_{3}\tau}{N_{0}}\mathrm{Ci}(\omega_{1}\tau) - \frac{\kappa_{3}\tau}{N_{0}}\left[ \frac{\pi}{2}-\mathrm{Si}(\omega_{1}\tau)\right] + \\
&+ &
\frac{2\sigma_{3}}{N_{0}}\sqrt{\tau}\left[iC_{0,2}(\sqrt{\omega_{1}\tau})-S_{0,2}(\sqrt{\omega_{1}\tau})\right].
\end{eqnarray}

\noindent The contrast and phase are readily found as

\begin{eqnarray}
\label{eq:9d}
\mu^{\mathrm{3D}}_{\mathrm{Hyb}}(\tau) &\approx& e^{-p_{e}\kappa_{3}\tau\left[ \frac{\pi}{2}-\mathrm{Si}(\omega_{1}\tau)\right] - 2 p_{e}\sigma_{3}\sqrt{\tau}S_{0,2}(\sqrt{\omega_{1}\tau})} \;\;\;\;\;
\end{eqnarray}
and
\begin{eqnarray}
\label{eq:9d}
\nu^{\mathrm{3D}}_{\mathrm{Hyb}}(\tau) &\approx& p_{e}\kappa_{3}\tau \mathrm{Ci}(\omega_{1}\tau) + 2p_{e}\sigma_{3}\sqrt{\tau}C_{0,2}(\sqrt{\omega_{1}\tau}). \;\;\;\;\;
\end{eqnarray}

This result shows that the contrast and phase signal for the hybrid potential are changing from the dipole-dipole solutions at an early delay to the van der Waals solutions at longer delay.
Also, by choosing $\omega_{1}$ sufficiently small the solution for the hybrid potential can be used as a substitute for the solution for the dipole-dipole interaction in three dimensions, where for the hybrid model the phase is clearly defined. Solutions for one and two dimensions can be obtained analogously.

\subsection{Averaging over the dipole trap}
For a sufficiently large and arbitrary shaped ensemble of atoms that locally exhibits a homogenous density distribution we can obtain the Ramsey signal of the whole ensemble by averaging over contributions coming from the individual atoms [see Eq.\eqref{pop}], which in the continuous limit becomes
\begin{eqnarray}
\label{eq:20y}
P(\tau) = \frac{1}{N}\int_{V} dV n({\bf R})P(n({\bf R}),\tau).
\end{eqnarray}
Here, $P(n({\bf R}),\tau)$ is equivalent to the solution found by the continuum approximation presented in this paragraph.
For example, for a Gaussian atom distribution we obtain
\begin{eqnarray}
\label{eq:21y}
P(\tau) &=& \frac{2}{\sqrt{\pi}n_{p}}\int_{0}^{n_{p}} dn\sqrt{\ln\left(\frac{n_{p}}{n}\right)}P(n,\tau),
\end{eqnarray}
where $n_{p}$ is the peak density.\\

\section{Conclusions}

We have theoretically investigated Ramsey dynamics of an interacting Rydberg gas for the purpose of identifying the mechanism which leads to experimentally observed decay of signal contrast and phase accumulation~\cite{Takei2015}. Our analysis provides a full formal solution to the Ramsey signal on which a novel approach involving a transition to the continuum is based. The analytical results obtained allow us to derive interesting scaling laws of direct experimental interest. We plan to extend our treatment in the future to describe interacting systems of quantum emitters undergoing coherent dynamics under diverse Hamiltonians (Ising, Heisenberg, etc). While in this paper we focused on the single-atom coherence, the results for the decay, phase shift, and the corresponding scaling laws can be applied to many-particle correlation functions which are presented in Appendix E.\\

\section{Acknowledgments}
This work was partially supported by CREST-JST, JSPS Grant-in-Aid for Specially Promoted Research Grant No. 16H06289 and for Young Scientists (B) Grant No. 15K17730, and Photon Frontier Network Program by MEXT.
G.P. was supported by ERC-St Grant ColdSIM (No. 307688), EOARD, RySQ, UdS via IdEX and ANR via BLUESHIELD.
K.O. thanks the Alexander von Humboldt foundation, University of Heidelberg, and University of Strasbourg for supporting this international collaboration. C.G. was supported from the Austrian Science Fund (FWF) via project No. P24968-N27.\\

\bibliography{scibib}

\begin{thebibliography}{99}


\bibitem{Bloch2008}
I. Bloch, J. Dalibard, W. Zwerger,
Many-body physics with ultracold gases.
{\it Rev. Mod. Phys.} {\bf 80}, 885--964 (2008).

\bibitem{Weimer2010}
H. Weimer, M. M\"uller, I. Lesanovsky, P. Zoller, H. P. B\"uchler,
A Rydberg quantum simulator.
{\it Nature Phys.} {\bf 6}, 382--388 (2010).

\bibitem{Martin2013}
M. J. Martin {\it et al.},
A quantum many-body spin system in an optical lattice clock.
{\it Science} {\bf 341}, 632--636 (2013).

\bibitem{Bettelli2013}
S. Bettelli {\it et al.},
Exciton dynamics in emergent Rydberg lattices.
{\it Phys. Rev. A} {\bf 88}, 043436 (2013).

\bibitem{Monroe2014}
P. Richerme {\it et al.},
Non-local propagation of correlations in quantum systems with long-range interactions.
{\it Nature} {\bf 511}, 198--201 (2014).

\bibitem{Jurcevic2014}
P. Jurcevic {\it et al.},
Quasiparticle engineering and entanglement propagation in a quantum many-body system.
{\it Nature} {\bf 511}, 202--205 (2014).

\bibitem{Yan2013}
B. Yan {\it et al.},
Observation of dipolar spin-exchange interactions with lattice-confined polar molecules.
{\it Nature} {\bf 501}, 521--525 (2013).

\bibitem{Schauss2012}
P. Schau\ss\ {\it et al.},
Observation of spatially ordered structures in a two-dimensional Rydberg gas.
{\it Nature} {\bf 491}, 87--91 (2012).

\bibitem{Schauss2015}
P. Schau\ss\ {\it et al.},
Crystallization in Ising quantum magnets.
{\it Science} {\bf 347}, 1455--1458 (2015).

\bibitem{Zeiher2016}
J. Zeiher {\it et al.},
Many-body interferometry of a Rydberg-dressed spin lattice.
{\it Nature Phys.} doi:10.1038/nphys3835 (2016)

\bibitem{Ramsey1950}
N. F. Ramsey,
A Molecular Beam Resonance Method with Separated Oscillating Fields.
{\it Phys. Rev.} {\bf 78}, 695--699 (1950).

\bibitem{Nipper2012}
J. Nipper {\it et al.},
Atomic pair-state interferometer: controlling and measuring an interaction-induced phase shift in Rydberg-atom pairs.
{\it Phys. Rev. X} {\bf 2}, 031011 (2012).

\bibitem{Alber1991}
G. Alber, P. Zoller,
Laser excitation of electronic wave packets in Rydberg atoms.
{\it Phys. Rep.} {\bf 199}, 231--280 (1991).

\bibitem{Noordam1992}
L. D. Noordam, D. I. Duncan, T. F. Gallagher,
Ramsey fringes in atomic Rydberg wave packets.
{\it Phys. Rev. A} {\bf 45}, 4734--4737 (1992).

\bibitem{Ohmori2009}
K. Ohmori,
Wave-packet and coherent control dynamics.
{\it Annu. Rev. Phys. Chem.} {\bf 60}, 487--511 (2009).

\bibitem{Kozak2013}
M. Koz\'ak, J. Precl\'ikov\'a, D. Fregenal, J. P. Hansen,
State-selective Rydberg excitation with femtosecond pulses.
{\it Phys. Rev. A} {\bf 87}, 043421 (2013).

\bibitem{Zhou2014}
T. Zhou, S. Li, R. R. Jones,
Rydberg-wave-packet evolution in a frozen gas of dipole-dipole-coupled atoms.
{\it Phys. Rev. A} {\bf 89}, 063413 (2014).

\bibitem{Takei2015}
N. Takei {\it et al.},
Direct observation of ultrafast many-body electron dynamics in a strongly-correlated ultracold Rydberg gas.
arXiv:1504.03635 (2015).

\bibitem{Worm2013}
M. van den Worm {\it et al.},
Relaxation timescales and decay of correlations in a long-range interacting quantum simulator.
{\it New J. Phys.} {\bf 15}, 083007 (2013).

\bibitem{Feig2013}
M. Foss-Feig {\it et al.},
Dynamical quantum correlations of Ising models on an arbitrary lattice and their resilience to decoherence.
{\it New J. Phys} {\bf 15}, 113008 (2013).

\bibitem{Hazzard2013}
K. R. A. Hazzard, S. R. Manmana, M. Foss-Feig, A. M. Rey,
Far-from-equilibrium Quantum Magnetism with Ultracold Polar Molecules.
{\it Phys. Rev. Lett.} {\bf 110}, 075301 (2013).

\bibitem{Hazzard2014}
K. R. A. Hazzard {\it et al.},
Quantum correlations and entanglement in far-from-equilibrium spin systems.
{\it Phys. Rev. A} {\bf 90}, 063622 (2014).

\bibitem{Mukherjee2015}
R. Mukherjee, T. C. Killian, K. R. A. Hazzard,
Accessing Rydberg-dressed interactions using many-body Ramsey dynamics.
arXiv:1511.08856 (2015).

\bibitem{Yeazell1988}
J. A. Yeazell, C. R. Stroud,
Observation of Spatially Localized Atomic Electron Wave Packets.
{\it Phys. Rev. Lett.} {\bf 60}, 1494 (1988).

\bibitem{tenWolde1988}
A. ten Wolde {\it et al.},
Observation of Radially Localized Atomic Electron Wave Packets.
{\it Phys. Rev. Lett.} {\bf 61}, 2099 (1988).

\bibitem{Yeazell1989}
J. A. Yeazell {\it et al.},
Classical periodic motion of atomic-electron wave packets.
{\it Phys. Rev. A} {\bf 40}, 5040 (1989).

\bibitem{Gould2004}
D. Tong {\it et al.},
Local blockade of Rydberg excitation in an ultracold gas.
{\it Phys. Rev. Lett.} {\bf 93}, 063001 (2004).

\bibitem{Urban2009}
E. Urban {\it et al.},
Observation of Rydberg blockade between two atoms.
{\it Nature Phys.} {\bf 5}, 110--114 (2009).

\bibitem{Gaetan2009}
A. Ga\"etan {\it et al.},
Observation of collective excitation of two individual atoms in the Rydberg blockade regime.
{\it Nature Phys.} {\bf 5}, 115--118 (2009).

\bibitem{Ates2007}
C. Ates, T. Pohl, T. Pattard, J. M. Rost,
Antiblockade in Rydberg Excitation of an Ultracold Lattice Gas.
{\it Phys. Rev. Lett.} {\bf 98}, 023002 (2007).

\bibitem{Amthor2013}
T. Amthor {\it et al.},
Evidence of Antiblockade in an Ultracold Rydberg Gas.
{\it Phys. Rev. Lett.} {\bf 104}, 013001 (2010).

\bibitem{Beguin2013}
L. B\'eguin, A. Vernier, R. Chicireanu, T. Lahaye, A. Browaeys,
Direct measurement of the van der Waals interaction Between Two Rydberg atoms.
{\it Phys. Rev. Lett.} {\bf 110}, 263201 (2013).

\bibitem{Ravets2014}
S. Ravets {\it et al.},
Coherent dipole-dipole coupling between two single Rydberg atoms at an electrically tuned F\"orster resonance.
{\it Nature Phys.} {\bf 10}, 914--917 (2014).

\bibitem{Zeiher2015}
J. Zeiher {\it et al.},
Microscopic Characterization of Scalable Coherent Rydberg Superatoms.
{\it Phys. Rev. X} {\bf 5}, 031015 (2015).

\bibitem{Saffman2016}
M. Saffman,
Quantum computing with atomic qubits and Rydberg interactions:  Progress and challenges.
{\it J.Phys. B: At. Mol. Opt. Phys.} {\bf 49}, 202001 (2016).

\bibitem{Saffman2013}
M. J. Piotrowicz, M. Lichtman, K. Maller, G. Li, S. Zhang, L. Isenhower, and M. Saffman,
Two-dimensional lattice of blue-detuned atom traps using a projected Gaussian beam array.
{\it Phys. Rev. A} {\bf 88}, 013420 (2013).

\bibitem{Browaeys2016}
 H. Labuhn, D. Barredo, S. Ravets, S. de L�s�leuc, T. Macr�, T. Lahaye, and A. Browaeys,
Realizing quantum Ising models in tunable two-dimensional arrays of single Rydberg atoms.
{\it  Nature} {\bf 534}, 667 (2016).

\bibitem{Whitlock2016a}
Y. Wang {\it et al.},
Magnetic lattices for ultracold atoms and degenerate quantum gases.
{\it Science Bulletin} {\bf 61}, 1097-1106 (2016).

\bibitem{Whitlock2016b}
S. Whitlock, A. W. Glaetzle, P. Hannaford,
Simulating Quantum Spin Models using Rydberg-Excited Atomic Ensembles in Magnetic Microtrap Arrays.
arXiv:1608.00251 (2016).



\bibitem{Hazzard2014a}
K. R. A. Hazzard {\it et al.},
Many-body dynamics of dipolar molecules in an optical lattice.
{\it Phys. Rev. Lett.} {\bf 113}, 195302 (2014).

\bibitem{Grimm2000}
R. Grimm, M. Weidem\"uller, Y. B. Ovchinnikov,
Optical dipole traps for neutral atoms.
{\it Advances in Atomic, Molecular and Optical Physics} {\bf 42}, 95-170 (2000).

\bibitem{Mathar2012}
R. J. Mathar,
Series Expansion of Generalized Fresnel Integrals.
arXiv:1211.3963 (2012).












































\end{thebibliography}

\bibliographystyle{Science}



\newpage

\section{Appendix A: Details on Ramsey pulses }

The pump and probe laser pulses ${\bf E}(t)$ and ${\bf E}(t-\tau)$ in the time-domain Ramsey interferometry are produced with a beam splitter in an optical interferometer and originate from one common laser pulse. For a two-level system  as presented in the main text with a ground state $|g\rangle$ and an excited state $|e\rangle$ and by neglecting interactions between the atoms in the excited state $|e\rangle$ during the pulse excitation we can describe the effect of a pulse on the state population and coherence by a unitary transformation $A$.\\
\indent The transformation can be obtained by solving the time-dependent Schr\"odinger equation for the first pulse excitation ${\bf E}(t)$
\begin{eqnarray}
\label{appa:eq1}
\nonumber
i\frac{\partial}{\partial t}|\psi\rangle = \left(\frac{\omega}{2}\hat{\sigma}_{z} -{\bf d}_{eg}\cdot {\bf E}(t)\hat{\sigma}_{+} -{\bf d}_{eg}^{*}\cdot {\bf E}(t)\hat{\sigma}_{-}\right)|\psi\rangle,\\
\end{eqnarray}
where $|\psi\rangle = C_{g}(t)|g\rangle + C_{e}(t)|e\rangle$ is the atomic state, ${\bf d}_{eg} = \langle e|{\bf d}|g\rangle $ is the transition dipole-matrix element, and ${\bf E}(t) = {\bf E}_{0}(\Theta(t-t_{0})-\Theta(t-t_{1}))\cos(\omega_{l}(t-t_{0}))$ a rectangular electric pulse with $\omega_{l}$ and $\Theta(t)$ being the laser frequency and the Heaviside step function, respectively.
Finally, the evolution by the pulse excitation is described by
\begin{eqnarray}
\label{appa:eq2}
\left(\begin{array}{c} C_{g}(t_{1}) \\ C_{e}(t_{1}) \end{array} \right) = \left( \begin{array}{cc} c_{g} & ic_{e} \\ ic_{e}^{*} & c_{g}^{*} \end{array} \right)\left( \begin{array}{c} C_{g}(t_{0}) \\ C_{e}(t_{0}) \end{array}\right).
\end{eqnarray}
Here, $c_{g} = \left(\cos(\Omega\delta t/2) - i\frac{\Delta}{\Omega}\sin(\Omega\delta t/2)\right)e^{i\omega_{l}\delta t/2}$ and $c_{e} = -\frac{\Omega_{R}}{\Omega}\sin(\Omega\delta t/2)e^{i\chi}e^{i\omega_{l}\delta t/2}$ with $\Omega = \sqrt{\Omega_{R}^{2} + \Delta^{2}}$, $\Delta = \omega-\omega_{l}$ being the detuning, $\tilde{\Omega}_{R} = \Omega_{R}e^{i\chi} = -{\bf d}_{eg}^{*}\cdot{\bf E}_{0}e^{-i\omega t_{0}}$ is the Rabi frequency, and $\delta t = t_{1} - t_{0}$ is the pulse duration.\\
\indent Solving the Schr\"odinger equation for the delayed pulse ${\bf E}(t-\tau)$ we obtain exactly the same matrix $A$ describing the state transformation.\\
\indent Using these definitions one can show that the phase $\phi$ first presented in the Ramsey signal of Eq.~\eqref{eq:Ram2} results from the ac-Stark shifted time evolution during the pulse excitation.

\section{Appendix B: Derivation of the general expression of the Ramsey signal}
In the following paragraph we describe the derivation of the time-domain Ramsey signal for an interacting many-particle system following an Ising type Hamiltonian $H$ as presented in Eq.~\eqref{Eq.2}. With all atoms being initially in the ground state $|g\rangle$ we obtain the state $|\Phi(\tau)\rangle_{N} = \exp(-iH\tau)A^{\otimes N}|g\rangle^{\otimes N}$  after a single pulse excitation, which transfers population to the strongly interacting excited state $|e\rangle$. The pumping is described by the $N$-particle tensor product of the matrix $A$. This is a viable description in the case the excitation happens on a much faster time scale than the interaction.\\
\indent Using an N-particle basis set which is defined by $|j_{1}\dots j_{N}\rangle$ with $j_{l} \in \{1,2\}, l\in \tilde{N} = \{1,\dots,N\}$, where $|g\rangle = |1\rangle$ and $|e\rangle = |2\rangle$ we can describe $|\Phi(\tau)\rangle_{N}$ by
\begin{eqnarray}
\label{ABEq.2}
\nonumber
|\Phi(\tau)\rangle_{N} = \sum_{j_{1},\dots,j_{N}} e^{-iH_{j_{1},\dots,j_{N}}\tau}A_{j_{1}1}\dots A_{j_{N}1}|j_{1}\dots j_{N}\rangle.\\
\end{eqnarray}
Here, $H_{j_{1},\dots,j_{N}} = \langle j_{N}\dots j_{1}|H|j_{1}\dots j_{N}\rangle$ can be written as
\begin{eqnarray}
\label{ABEq.3}
\nonumber
H_{j_{1},\dots,j_{N}} &=& \left( -\sum_{l=1}^{N}\frac{\omega_{l}}{2}\right) + \sum_{l\in I^{\bf j}_{N}}\omega_{l} + \sum_{l\in I^{\bf j}_{N}}\sum_{k\in  I^{\bf j}_{N}}\frac{U_{lk}}{2} \\
&=& \eta_{N} + \sum_{l\in I^{\bf j}_{N}}\omega_{l} + \sum_{l\in I^{\bf j}_{N}}\sum_{k\in  I^{\bf j}_{N}}\frac{U_{lk}}{2},
\end{eqnarray}
with $I^{\bf j}_{N} = \{ l\in \tilde{N}|  j_{l} = 2 \}$, ${\bf j} = (j_{1},\dots,j_{N})$ and $\eta_{N}= \left( -\sum_{l=1}^{N}\frac{\omega_{l}}{2}\right)$.
This allows us to write down the density operator for the pure $N$-particle state
\begin{widetext}
\begin{eqnarray}
\label{ABEq.4}
\hat{\rho}_{1\dots N} &=& |\Phi(\tau)\rangle_{N}\langle \Phi(\tau)|_{N} \\\nonumber
&=& \sum_{j_{1}\dots j_{N} k_{1}\dots k_{N}} e^{-i\left( H_{j_{1},\dots,j_{N}}-H_{k_{1},\dots,k_{N}}\right)\tau}\left( A_{j_{1}1}A^{*}_{k_{1}1}\right)\dots \left(A_{j_{N}1}A^{*}_{k_{N}1}\right)|j_{1}\dots j_{N}\rangle \langle k_{N}\dots k_{1}|.
\end{eqnarray}
\end{widetext}
The reduced density operator for $N-1$ particles can be obtained by taking the trace over the terms from the $N$'th particle, where we choose $N$ without loss of generality. This is given by
\begin{widetext}
\begin{eqnarray}
\label{ABEq.5}
\hat{\rho}_{1\dots N-1} &=& \sum_{j_{1}\dots j_{N-1} k_{1}\dots k_{N-1}} \left( e^{-i\left( H_{j_{1},\dots,j_{N-1},1}-H_{k_{1},\dots,k_{N-1},1}\right)\tau}|A_{11}|^{2} \right. \\\nonumber
& & \left. + e^{-i\left( H_{j_{1},\dots,j_{N-1},2}-H_{k_{1},\dots,k_{N-1},2}\right)\tau}|A_{21}|^{2}\right)\left( A_{j_{1}1}A^{*}_{k_{1}1}\right)\dots \left(A_{j_{N-1}1}A^{*}_{k_{N-1}1}\right)|j_{1}\dots j_{N-1}\rangle \langle k_{N-1}\dots k_{1}|.
\end{eqnarray}
\end{widetext}
Since
\begin{eqnarray}
\label{ABEq.6}
\nonumber
H_{j_{1},\dots,j_{N-1},1}-H_{k_{1},\dots,k_{N-1},1} =  H_{j_{1},\dots,j_{N-1}}-H_{k_{1},\dots,k_{N-1}}\\
\end{eqnarray}
and
\begin{eqnarray}
\label{ABEq.7}
H_{j_{1},\dots,j_{N-1},2}&-&H_{k_{1},\dots,k_{N-1},2} = H_{j_{1},\dots,j_{N-1}}-\\\nonumber
& & -H_{k_{1},\dots,k_{N-1}} + \sum_{a\in I^{\bf j}_{N-1}} U_{aN} - \sum_{c\in I^{\bf k}_{N-1}} U_{cN},
\end{eqnarray}
we obtain for the reduced density operator
\begin{widetext}
\begin{eqnarray}
\label{ABEq.8}
\hat{\rho}_{1\dots N-1} &=& \sum_{j_{1}\dots j_{N-1} k_{1}\dots k_{N-1}} \left(p_{g} + e^{-i\left(\sum_{a\in I^{\bf j}_{N-1}} U_{aN} -  \sum_{c\in I^{\bf k}_{N-1}} U_{cN}\right)\tau}p_{e}\right)\\\nonumber
& & \times e^{-i\left( H_{j_{1},\dots,j_{N-1}}-H_{k_{1},\dots,k_{N-1}}\right)\tau}\left( A_{j_{1}1}A^{*}_{k_{1}1}\right)\dots \left(A_{j_{N-1}1}A^{*}_{k_{N-1}1}\right)|j_{1}\dots j_{N-1}\rangle \langle k_{N-1}\dots k_{1}|.
\end{eqnarray}
\end{widetext}
This procedure can be continued for other particles. In the limit we obtain the reduced density operator for single particles in the interacting $N$-particle ensemble. Here, without loss of generality the reduced density operator of the first particle is given by
\begin{widetext}
\begin{eqnarray}
\label{ABEq.10}
\hat{\rho}_{1} &=& \sum_{j_{1} k_{1}} \prod_{l = 2}^{N}\left(p_{g} + e^{-i\left(\sum_{a\in I^{\bf j}_{1}} U_{al} -  \sum_{c\in I^{\bf k}_{1}} U_{cl}\right)\tau}p_{e}\right)e^{-i\left( H_{j_{1}}-H_{k_{1}}\right)\tau} \left( A_{j_{1}1}A^{*}_{k_{1}1}\right)|j_{1}\rangle \langle k_{1}|\\\nonumber
&=& |c_{g}|^{2}|g\rangle\langle g| - i\prod_{l = 2}^{N}\left(p_{g} + e^{iU_{1l}\tau}p_{e}\right) e^{i\omega\tau}c_{g}c_{e}|g\rangle\langle e| + i\prod_{l = 2}^{N}\left(p_{g} + e^{-iU_{1l}\tau}p_{e}\right) e^{-i\omega\tau}c^{*}_{e}c^{*}_{g}|e\rangle\langle g| + |c_{e}|^{2}|e\rangle\langle e|,
\end{eqnarray}
\end{widetext}
for $j=1$. This density operator gives the populations and coherences of a mixed state if $\tau \neq 0$. The second pulse which arrives at the delay time $\tau$ rotates the signal of the non diagonal terms which represent the coherence of a single particle interacting with $N-1$ surrounding atoms as shown in Eq.~\eqref{ABEq.10} into the population of the ground- and Rydberg-state atoms. When we define $G(\tau) = \prod_{l=2}^{N}\left( p_{g} + p_{e}e^{iU_{1l}\tau}\right)$ the double-pulse time-domain Ramsey process can be described by
\begin{widetext}
\begin{eqnarray}
\label{ABEq.10a}
\hat{\rho}^{2\mathrm{p}}_{1} &=& A\hat{\rho}_{1}A^{*}\\\nonumber
&=& \left(1 - 2p_{g}p_{e}\Re\left[ 1+ e^{i(\omega\tau+\phi)}G(\tau)\right]    \right)|g\rangle\langle g| - i c_{g}c_{e}\left(p_{g}\left(1 + e^{i(\omega\tau+\phi)}G(\tau)\right)-p_{e}\left(1 +e^{-i(\omega\tau+\phi)}G^{*}(\tau)\right)\right)|g\rangle\langle e|\\\nonumber
& &+ic^{*}_{g}c^{*}_{e}\left(p_{g}\left(1 + e^{-i(\omega\tau+\phi)}G^{*}(\tau)\right)-p_{e}\left(1 +e^{i(\omega\tau+\phi)}G(\tau)\right)\right)|e\rangle\langle g|+\left(2p_{g}p_{e}\Re\left[ 1+ e^{i(\omega\tau+\phi)}G(\tau)\right]\right)|e\rangle\langle e|.
\end{eqnarray}
\end{widetext}
By evaluating the population in the excited state $|e\rangle$ after a Ramsey experiment we obtain the signal
\begin{eqnarray}
\label{ABEq.11}
P(\tau) &=& 2p_{g}p_{e}\Re\left[ 1+ e^{i(\omega\tau+\phi)}G(\tau)\right]\\\nonumber
 &=& 2p_{g}p_{e}\Re\left[ 1+ e^{i(\omega\tau+\phi)}\prod_{l=2}^{N}\left( p_{g} + p_{e}e^{iU_{1l}\tau}\right)\right].
\end{eqnarray}

\section{Appendix C: Details on the discrete to continuum transformation}

For a large atom number $N$ the sums in Eq.~\eqref{eq:Approx2} can be substituted by integrals and we obtain oscillation terms given by

\begin{widetext}
\begin{eqnarray}
\label{eq:Approx3}
B_{1} &=& \sum_{i_{1}=2}^{N} e^{iU(r_{1,i_{1}})\tau} = 4\pi\int r^{2}dr f_{1}(r)e^{i\ U(r)\tau}, \\\nonumber
B_{2} = \sum_{i_{1}=2}^{N-1}\sum_{i_{2} > i_{1}}^{N} e^{i(U(r_{1,i_{1}})+ U(r_{1,i_{2}}))\tau} &=&
(4\pi)^{2}\int \int r^{2}_{1}dr_{1} r^{2}_{2}dr_{2} f_{2}(r_{1},r_{2})e^{i\left(U(r_{1}) + U(r_{2})\right)\tau},\\\nonumber
B_{3} &=& \cdots.
\end{eqnarray}
\end{widetext}
The functions $f_{m}$ are contrained by
\begin{eqnarray}
\label{eq:Approx4}
\int_{V_{1}} \cdots \int_{V_{m}} dV_{1} \cdots dV_{m} f_{m}  = \left( \begin{array}{c} N-1 \\ m \end{array} \right).
\end{eqnarray}

For an ensemble with a large number of atoms $N$ which is locally described by a homogenous atom distribution we can approximate the function $f_{m}$ with a constant value $f_{m}  \approx (1/V_{d}(r_{0}))^{m} \binom{N_{0}-1}{m}$ where $V_{d}(r_{0}) = \frac{N_{0}}{n}= \frac{\pi^{d/2}r_{0}^{d}}{\Gamma(d/2+1)}$ is a local spherical volume in $d$-dimensions that encompasses $N_{0}$ atoms.

Using the identity $p_{g} + p_{e} = 1$ and the convergence of the Ramsey signal for distances $r \geq r_{0}$ obtained by incorporating large enough atom numbers $N_{0}$ allowing us to ignore interactions beyond $r_{0}$ we can substitute $N$ by $N_0$, which leads to
\begin{eqnarray}
\label{eq:Approx5}
G(\tau) &\approx& \left( p_{g} + p_{e}\gamma(\tau)\right)^{N_{0}-1},
\end{eqnarray}
where we ignore the index of the atom in the following text of the paragraph. We use $\gamma(\tau)=\frac{d}{r_{0}^{d}-r_{\mathrm{B}}^{d}} I(\tau)$ where $I(\tau)=\int_{r_{\mathrm{B}}}^{r_{0}}dr r^{d-1} e^{iU(r)\tau}$. The radius $r_{\mathrm{B}}$ can be used to describe the restriction due to the limited bandwidth of the pump and probe excitation and is equivalent to the Rydberg-blockade radius.
In the case of a van der Waals and dipole-dipole interaction described by $U(r) = C_{6}/r^{6}$ and $U(r)=C_{3}/r^{3}$ we can simplify this function $\gamma(\tau)$ even further by evaluating the integrals:
\begin{widetext}
\begin{eqnarray}
\label{eq:2}
I^{d\mathrm{D}}_6(\tau) =  \frac{\sqrt[6/d]{C_{6}\tau}}{d}\left ( \left [ -\frac{e^{is}}{\sqrt[6/d]{s}} \right ]_{s_{0}}^{s_{\mathrm{B}}}+ i\left(\frac{6}{d}\right)(\tilde{C}_{d}(x_{\mathrm{B}})-\tilde{C}_{d}(x_{0})) - 2(\tilde{S}_{d}(x_{\mathrm{B}})-\tilde{S}_{d}(x_{0})) \right)
\end{eqnarray}
\end{widetext}
and
\begin{widetext}
\begin{eqnarray}
\label{eq:1aa}
I^{1\mathrm{D}}_{3}(\tau) &=& C^{1/3}_{3}\left(\left[\frac{e^{i\omega\tau}}{\omega^{1/3}} \right]^{\omega_{0}}_{\omega_{\mathrm{B}}} + 3\tau^{\frac{1}{3}}\left(i(C_{1,3}(\sqrt[3]{\omega_{\mathrm{B}}\tau})-C_{1,3}(\sqrt[3]{\omega_{0}\tau}) )-(S_{1,3}(\sqrt[3]{\omega_{\mathrm{B}}\tau})-S_{1,3}(\sqrt[3]{\omega_{0}\tau}) \right) )\right) \\\nonumber
I^{2\mathrm{D}}_{3}(\tau) &=& \frac{C^{2/3}_{3}}{2} \left(\left[\frac{e^{i\omega\tau}}{\omega^{2/3}} \right]^{\omega_{0}}_{\omega_{\mathrm{B}}} + i3\tau\left[\sqrt[3]{\omega}e^{i\omega\tau}\right]^{\omega_{\mathrm{B}}}_{\omega_{0}} + 9\tau^{\frac{2}{3}}\left((C_{3,3}(\sqrt[3]{\omega_{\mathrm{B}}\tau})-C_{3,3}(\sqrt[3]{\omega_{0}\tau}) )+i(S_{3,3}(\sqrt[3]{\omega_{\mathrm{B}}\tau})-S_{3,3}(\sqrt[3]{\omega_{0}\tau}) \right) )\right) \\
\\
I^{3\mathrm{D}}_3(\tau) &=& \frac{C_{3}}{3}\left ( \left [ -\frac{e^{i\omega\tau}}{\omega} \right ]_{\omega_{0}}^{\omega_{\mathrm{B}}}+ i\tau\left(\mathrm{Ci}(\omega_{\mathrm{B}}\tau) - \mathrm{Ci}(\omega_{0}\tau) \right)  - \tau\left(\mathrm{Si}(\omega_{\mathrm{B}}\tau)-\mathrm{Si}(\omega_{0}\tau) \right) \right).
\end{eqnarray}
\end{widetext}
For the van der Waals interaction with $\omega = C_{6}/r^{6}$ we have used $s= \omega\tau$, $x = \sqrt[6/d]{s}$ and the Fresnel integrals $\tilde{S}_{d}(x) = S_{m=(3-d)^{2}, n = 6/d}(x)= \int_{0}^{x}t^{m}\sin(t^{n})dt$  and $\tilde{C}_{d}(x) = C_{m=(3-d)^{2}, n=6/d}(x) = \int_{0}^{x}t^{m}\cos(t^{n})dt$.
For the dipole-dipole interaction we have used $\omega = C_{3}/r^{3}$, the Fresnel integrals $C_{1,3}$, $C_{3,3}$, $S_{1,3}$, $S_{3,3}$ and $\mathrm{Si}(x) = \int_{0}^{x}dt\sin(t)/t$ and $\mathrm{Ci}(x) = -\int_{x}^{\infty}dt\cos(t)/t$.

For the prefactor of the integral we derive
\begin{eqnarray}
\label{eq:4}
\frac{d}{r_{0}^{d}-r_{\mathrm{B}}^{d}} &=& d\sqrt[6/d]{\frac{1}{C_{6}}}\left (\frac{\sqrt[6/d]{\omega_{\mathrm{B}}\omega_{0}}}{\sqrt[6/d]{\omega_{\mathrm{B}}}-\sqrt[6/d]{\omega_{0}}} \right) \;\mathrm{vdW}\\
\frac{d}{r_{0}^{d}-r_{\mathrm{B}}^{d}} &=& d\sqrt[3/d]{\frac{1}{C_{3}}}\left (\frac{\sqrt[3/d]{\omega_{\mathrm{B}}\omega_{0}}}{\sqrt[3/d]{\omega_{\mathrm{B}}}-\sqrt[3/d]{\omega_{0}}} \right) \;\;\mathrm{DD}.
\end{eqnarray}
This leads to
\begin{widetext}
\begin{eqnarray}
\label{eq:5}
\nonumber
\gamma^{d\mathrm{D}}_{\mathrm{vdW}}(\tau) &=& \left (\frac{\sqrt[6/d]{\omega_{\mathrm{B}}}e^{i\omega_{0}\tau}-\sqrt[6/d]{\omega_{0}}e^{i\omega_{\mathrm{B}}\tau}}{\sqrt[6/d]{\omega}_{\mathrm{B}}-\sqrt[6/d]{\omega}_{0}}\right)\\
& &+ \left (\frac{\left(\frac{6}{d}\right)\sqrt[6/d]{\omega_{\mathrm{B}}\omega_{0}\tau}}{\sqrt[6/d]{\omega_{\mathrm{B}}}-\sqrt[6/d]{\omega_{0}}} \right) \left(i\left(\tilde{C}_{d}(\sqrt[6/d]{\omega_{\mathrm{B}}\tau}) - \tilde{C}_{d}(\sqrt[6/d]{\omega_{0}\tau}) \right)  - \left(\tilde{S}_{d}(\sqrt[6/d]{\omega_{\mathrm{B}}\tau})-\tilde{S}_{d}(\sqrt[6/d]{\omega_{0}\tau})\right)\right)\\
\gamma^{1\mathrm{D}}_{\mathrm{DD}}(\tau) &=& \gamma^{2\mathrm{D}}_{\mathrm{vdW}} \\
\gamma^{2\mathrm{D}}_{\mathrm{DD}}(\tau) &=& \left (\frac{\omega_{\mathrm{B}}^{2/3}e^{i\omega_{0}\tau}-\omega_{0}^{2/3}e^{i\omega_{\mathrm{B}}\tau}}{\omega^{2/3}_{\mathrm{B}}-\omega^{2/3}_{0}}\right) -i3\tau\left (\frac{\omega_{\mathrm{B}}^{2/3}\omega_{0}e^{i\omega_{0}\tau}-\omega_{0}^{2/3}\omega_{\mathrm{B}}e^{i\omega_{\mathrm{B}}\tau}}{\omega^{2/3}_{\mathrm{B}}-\omega^{2/3}_{0}}\right)\\
& &+ \frac{9\tau^{2/3}(\omega_{0}\omega_{\mathrm{B}})^{2/3}}{\left(\omega^{2/3}_{\mathrm{B}} - \omega^{2/3}_{0} \right)}\left((C_{3,3}(\sqrt[3]{\omega_{\mathrm{B}}\tau})-C_{3,3}(\sqrt[3]{\omega_{0}\tau}) )+i(S_{3,3}(\sqrt[3]{\omega_{\mathrm{B}}\tau})-S_{3,3}(\sqrt[3]{\omega_{0}\tau}) \right) ) \\
\gamma^{3\mathrm{D}}_{\mathrm{DD}}(\tau) &=& \left (\frac{\omega_{\mathrm{B}}e^{i\omega_{0}\tau}-\omega_{0}e^{i\omega_{\mathrm{B}}\tau}}{\omega_{\mathrm{B}}-\omega_{0}}\right) + \left (\frac{\omega_{\mathrm{B}}\omega_{0}\tau}{\omega_{\mathrm{B}}-\omega_{0}} \right) (i\left(\mathrm{Ci}(\omega_{\mathrm{B}}\tau) - \mathrm{Ci}(\omega_{0}\tau) \right)  - \left(\mathrm{Si}(\omega_{\mathrm{B}}\tau)-\mathrm{Si}(\omega_{0}\tau)\right)).
\end{eqnarray}
\end{widetext}
In contrast to $\omega_{\mathrm{B}}$ which is a constant number $\omega_{0}$ depends on the number of atoms $N_{0}$. This follows from the relations
$N_{0}/n = V_{d} = \frac{\pi^{d/2}}{\Gamma(d/2+1)}r_{0}^{d}$ and $r_{0} = (C_{k}/\omega_{0})^{1/k}$ resulting in
\begin{eqnarray}
\label{eq:6}
\omega_{0}^{d/6} &=& \frac{\pi^{d/2} n C_{6}^{d/6}}{\Gamma(d/2+1)}\frac{1}{N_{0}} = \sigma_{d}\frac{1}{N_{0}} \;\;\;\mathrm{vdW}\\
\omega_{0}^{d/3} &=& \frac{\pi^{d/2} n C_{3}^{d/3}}{\Gamma(d/2+1)}\frac{1}{N_{0}} = \kappa_{d}\frac{1}{N_{0}} \;\;\;\; \mathrm{DD}.
\end{eqnarray}
For increasing atom number $N_{0}$ and the condition $\omega^{d/6}_{\mathrm{B}} > \sigma_{d}$ and $\omega^{d/3}_{\mathrm{B}} > \kappa_{d}$, the beginning terms of $\gamma_{\mathrm{vdW}}(\tau)$ and $\gamma_{\mathrm{DD}}(\tau)$ can be approximated by
\begin{widetext}
\begin{eqnarray}
\label{eq:7}
 \left (\frac{\sqrt[6/d]{\omega_{\mathrm{B}}}e^{i\omega_{0}\tau}-\sqrt[6/d]{\omega_{0}}e^{i\omega_{\mathrm{B}}\tau}}{\sqrt[6/d]{\omega}_{\mathrm{B}}-\sqrt[6/d]{\omega}_{0}}\right) \approx 1 + \frac{\sigma_{d}\left(1 - \cos(\omega_{\mathrm{B}}\tau) - i\sin(\omega_{\mathrm{B}}\tau))\right)}{N_{0}\sqrt[6/d]{\omega_{\mathrm{B}}}}
\end{eqnarray}
\end{widetext}
 and
\begin{widetext}
\begin{eqnarray}
\label{eq:6}
 \left (\frac{\sqrt[3]{\omega_{\mathrm{B}}}e^{i\omega_{0}\tau}-\sqrt[3]{\omega_{0}}e^{i\omega_{\mathrm{B}}\tau}}{\sqrt[3]{\omega_{\mathrm{B}}}-\sqrt[3]{\omega_{0}}}\right) &\approx & 1 + \frac{\kappa_{1}(1 - \cos(\omega_{\mathrm{B}}\tau) - i\sin(\omega_{\mathrm{B}}\tau))}{N_{0}\sqrt[3]{\omega_{\mathrm{B}}}}\\
 \left (\frac{\omega^{2/3}_{\mathrm{B}}e^{i\omega_{0}\tau}-\omega^{2/3}_{0}e^{i\omega_{\mathrm{B}}\tau}}{\omega^{2/3}_{\mathrm{B}}-\omega^{2/3}_{0}}\right) -i3\tau  \left (\frac{\omega^{2/3}_{\mathrm{B}}\omega_{0}e^{i\omega_{0}\tau}-\omega^{2/3}_{0}\omega^{2/3}_{\mathrm{B}}e^{i\omega_{\mathrm{B}}\tau}}{\omega^{2/3}_{\mathrm{B}}-\omega^{2/3}_{0}}\right)  &\approx & 1 + \frac{\kappa_{2}(1 -(1-i3\omega_{\mathrm{B}}\tau)\exp(i\omega_{\mathrm{B}}\tau))}{N_{0}\omega^{2/3}_{\mathrm{B}}}\\
 \left (\frac{\omega_{\mathrm{B}}e^{i\omega_{0}\tau}-\omega_{0}e^{i\omega_{\mathrm{B}}\tau}}{\omega_{\mathrm{B}}-\omega_{0}}\right) &\approx & 1 + i\frac{\kappa_{3}\tau}{N_{0}} + \frac{\kappa_{3}(1 - \cos(\omega_{\mathrm{B}}\tau) - i\sin(\omega_{\mathrm{B}}\tau))}{N_{0}\omega_{\mathrm{B}}}.\;\;\;\;
\end{eqnarray}
\end{widetext}
Here, the prefactors of the last terms follow
\begin{eqnarray}
\label{eq:8a}
\left (\frac{\left(\frac{6}{d}\right)\sqrt[6/d]{\omega_{\mathrm{B}}\omega_{0}\tau}}{\sqrt[6/d]{\omega_{\mathrm{B}}}-\sqrt[6/d]{\omega_{0}}} \right)  &\approx& \left(\frac{6}{d}\right)\frac{\sigma_{d}}{N_{0}}\sqrt[6/d]{\tau}\\
\label{eq:8}
\left (\frac{(\omega_{\mathrm{B}}\omega_{0}\tau)^{d/3}}{\omega^{d/3}_{\mathrm{B}}-\omega^{d/3}_{0}} \right) &\approx & \frac{\kappa_{d}}{N_{0}}\tau^{d/3}
\end{eqnarray}
for large $N_{0}$.\\
\indent In the case of a van der Waals interaction we acquire the relation
\begin{widetext}
\begin{eqnarray}
\label{eq:9a}
\gamma^{d\mathrm{D}}_{\mathrm{vdW}}(\tau) &=& 1 + \frac{\sigma_{d}\left(1 - \cos(\omega_{\mathrm{B}}\tau) - i\sin(\omega_{\mathrm{B}}\tau))\right)}{N_{0}\sqrt[6/d]{\omega_{\mathrm{B}}}} + \dots \\\nonumber
& &+ \frac{6\sigma_{d}}{N_{0}d}\sqrt[6/d]{\tau}\left( i\left(\tilde{C}_{d}(\sqrt[6/d]{\omega_{\mathrm{B}}\tau}) - \tilde{C}_{d}((\sigma_{d}/N_{0})\sqrt[6/d]{\tau}) \right)  - \left(\tilde{S}_{d}(\sqrt[6/d]{\omega_{\mathrm{B}}\tau})-\tilde{S}_{d}((\sigma_{d}/N_{0})\sqrt[6/d]{\tau})\right)\right) \\\nonumber
 &\approx& 1 + \frac{\sigma_{d}\left(1 - \cos(\omega_{\mathrm{B}}\tau) - i\sin(\omega_{\mathrm{B}}\tau)\right)}{N_{0}\sqrt[6/d]{\omega_{\mathrm{B}}}} + \frac{6\sigma_{d}}{N_{0}d}\sqrt[6/d]{\tau}\left( i\left(\tilde{C}_{d}(\sqrt[6/d]{\omega_{\mathrm{B}}\tau}) \right)  - \left(\tilde{S}_{d}(\sqrt[6/d]{\omega_{\mathrm{B}}\tau})\right)\right).
\end{eqnarray}
\end{widetext}
For $\omega_{\mathrm{B}} \rightarrow \infty$ this results in
\begin{eqnarray}
\label{eq:9bx}
\gamma^{\mathrm{1D}}_{\mathrm{vdW}} &\approx& 1 + \frac{\sigma_{1}\Gamma(5/6)}{N_{0}}\sqrt[6]{\tau}\left(i\exp\left(\frac{i5\pi}{12}\right)\right)\;\;\;\;\; \\
\label{eq:9cx}
\gamma^{\mathrm{2D}}_{\mathrm{vdW}} &\approx& 1 + \frac{\sigma_{2}\Gamma(2/3)}{N_{0}}\sqrt[3]{\tau}\left(i\exp\left(\frac{i\pi}{3}\right)\right) \\
\label{eq:9bx}
\gamma^{\mathrm{3D}}_{\mathrm{vdW}} &\approx& 1 + \frac{2\sigma_{3}}{N_{0}}\sqrt{\frac{\pi}{8}}\sqrt{\tau}(i -1).
\end{eqnarray}
Similarly, for the dipole-dipole interaction we obtain
\begin{widetext}
\begin{eqnarray}
\label{eq:9xxx}
\nonumber
\gamma^{\mathrm{1D}}_{\mathrm{DD}}(\tau) &=& 1 + \frac{\kappa_{1}(1 - \cos(\omega_{\mathrm{B}}\tau) - i\sin(\omega_{\mathrm{B}}\tau))}{N_{0}\sqrt[3]{\omega_{\mathrm{B}}}}\\
& & + \frac{3\kappa_{1}}{N_{0}}\sqrt[3]{\tau}\left(i\left(C_{1,3}(\sqrt[3]{\omega_{\mathrm{B}}\tau}) - C_{1,3}((\kappa_{1}/N_{0})\sqrt[3]{\tau}) \right)  - \left(S_{1,3}(\sqrt[3]{\omega_{\mathrm{B}}\tau})-S_{1,3}((\kappa_{1}/N_{0})\sqrt[3]{\tau})\right)\right)\\\nonumber
&\approx & 1 + \frac{\kappa_{1}(1 - \cos(\omega_{\mathrm{B}}\tau) - i\sin(\omega_{\mathrm{B}}\tau))}{N_{0}\sqrt[3]{\omega_{\mathrm{B}}}} + \frac{3\kappa_{1}}{N_{0}}\sqrt[3]{\tau}\left(iC_{1,3}(\sqrt[3]{\omega_{\mathrm{B}}\tau}) - S_{1,3}(\sqrt[3]{\omega_{\mathrm{B}}\tau})\right) \\\nonumber
\gamma^{\mathrm{2D}}_{\mathrm{DD}}(\tau) &=& 1 + \frac{\kappa_{2}(1 -(1-i3\omega_{\mathrm{B}}\tau)\exp(i\omega_{\mathrm{B}}\tau))}{N_{0}\omega^{2/3}_{\mathrm{B}}}\\
& & + \frac{9\kappa_{2}}{N_{0}}\tau^{2/3} \left(\left(C_{3,3}(\sqrt[3]{\omega_{\mathrm{B}}\tau}) - C_{3,3}(\sqrt{\kappa_{2}/N_{0}}\sqrt[3]{\tau}) \right)  + i\left(S_{3,3}(\sqrt[3]{\omega_{\mathrm{B}}\tau})-S_{3,3}(\sqrt{\kappa_{2}/N_{0}}\sqrt[3]{\tau})\right)\right)\\\nonumber
&\approx & 1 + \frac{\kappa_{2}(1 -(1-i3\omega_{\mathrm{B}}\tau)\exp(i\omega_{\mathrm{B}}\tau))}{N_{0}\omega^{2/3}_{\mathrm{B}}} + \frac{9\kappa_{2}}{N_{0}}\tau^{2/3} \left(C_{3,3}(\sqrt[3]{\omega_{\mathrm{B}}\tau})  + iS_{3,3}(\sqrt[3]{\omega_{\mathrm{B}}\tau})\right) \\\nonumber
\gamma^{\mathrm{3D}}_{\mathrm{DD}}(\tau) &=& 1 + i\frac{\kappa_{3}\tau}{N_{0}} + \frac{\kappa_{3}(1 - \cos(\omega_{\mathrm{B}}\tau) - i\sin(\omega_{\mathrm{B}}\tau))}{N_{0}\omega_{\mathrm{B}}} + \dots \\
& & + \frac{\kappa_{3}}{N_{0}}\tau (i\left(\mathrm{Ci}(\omega_{\mathrm{B}}\tau) - \mathrm{Ci}((\kappa_{3}/N_{0})\tau) \right)  - \left(\mathrm{Si}(\omega_{\mathrm{B}}\tau)-\mathrm{Si}((\kappa_{3}/N_{0})\tau)\right))\\\nonumber
&\approx& 1 + i\frac{\kappa_{3}\tau}{N_{0}}+  \frac{\kappa_{3}(1 - \cos(\omega_{\mathrm{B}}\tau) - i\sin(\omega_{\mathrm{B}}\tau))}{N_{0}\omega_{\mathrm{B}}} + \frac{\kappa_{3}}{N_{0}}\tau (i(\mathrm{Ci}(\omega_{\mathrm{B}}\tau)-\mathrm{Ci}((\kappa_{3}/N_{0})\tau)) + \mathrm{Si}(\omega_{\mathrm{B}}\tau)).
\end{eqnarray}
\end{widetext}
Additionally, by using the approximation $\mathrm{Ci}(x) = \gamma + \ln(x) + \dots$ valid for small $x$ where $\gamma$ is the Euler-Mascheroni constant, we obtain for $\omega_{\mathrm{B}} \rightarrow \infty$  the approximations
\begin{eqnarray}
\label{eq:9z}
\gamma^{\mathrm{3D}}_{\mathrm{DD}} &\approx& 1 + i\frac{\kappa_{3}\tau}{N_{0}} - \frac{\kappa_{3}}{N_{0}}\tau (i(\gamma + \ln(\kappa_{3}\tau/N_{0})) + \pi/2).\;\;\;\;
\end{eqnarray}
The coherence terms $\tilde{G}(\tau) = \alpha(\tau)G(\tau)$ for the van der Waals interaction are easily obtained by taking the limit for large atom numbers $N_{0}$ given by
\begin{widetext}
\begin{eqnarray}
\label{eq:11}
 \tilde{G}^{d\mathrm{D}}_{\mathrm{vdW}}(\tau) &\approx& \alpha(\tau)\left( 1 +  \frac{p_{e}\sigma_{d}\Gamma\left(\frac{6-d}{6}\right)}{N_{0}}\sqrt[6/d]{\tau}\left(i\exp\left(\frac{i(6-d)\pi}{12}\right)\right)\right)^{N_{0}}\\
&\approx& e^{i\left(\omega\tau + \phi + p_{e}\sigma_{d}\Gamma\left(\frac{6-d}{6}\right)\cos\left(\frac{(6-d)\pi}{12}\right)\sqrt[6/d]{\tau}\right)}e^{-p_{e}\sigma_{d}\Gamma\left(\frac{6-d}{6}\right)\sin\left(\frac{(6-d)\pi}{12}\right)\sqrt[6/d]{\tau}},
\end{eqnarray}
\end{widetext}
which results in the Ramsey signal
\begin{eqnarray}
\label{eq:11a}
\nonumber
P(n,\tau) &=& 2p_{g}p_{e}\Re\left\{ 1+ e^{i\left(\omega\tau + \phi + p_{e}\sigma_{d}\Gamma\left(\frac{6-d}{6}\right)\cos\left(\frac{(6-d)\pi}{12}\right)\sqrt[6/d]{\tau}\right)} \right. \\
& & \times \left. e^{-p_{e}\sigma_{d}\Gamma\left(\frac{6-d}{6}\right)\sin\left(\frac{(6-d)\pi}{12}\right)\sqrt[6/d]{\tau}} \right\}.
\end{eqnarray}

The functions $\tilde{G}(\tau)$ for the dipole-dipole interaction are given by
\begin{widetext}
\begin{eqnarray}
\label{eq:9yxx}
\tilde{G}^{\mathrm{1D}}_{\mathrm{DD}}(\tau) &\approx& e^{i\left(\omega\tau + \phi + p_{e}\kappa_{1}\Gamma\left(\frac{2}{3}\right)\cos\left(\frac{\pi}{3}\right)\sqrt[3]{\tau}\right)}e^{-p_{e}\kappa_{1}\Gamma\left(\frac{2}{3}\right)\sin\left(\frac{\pi}{3}\right)\sqrt[3]{\tau}}\\\nonumber
\tilde{G}^{\mathrm{2D}}_{\mathrm{DD}}(\tau) &\approx& e^{i\left(\omega\tau + \phi - p_{e}\kappa_{2}\left(\frac{\sin(\omega_{\mathrm{B}}\tau)-3\omega_{\mathrm{B}}\tau\cos(\omega_{\mathrm{B}}\tau)}{\omega_{\mathrm{B}}^{2/3}} - 9\tau^{\frac{2}{3}}S_{3,3}\left(\sqrt[3]{\omega_{\mathrm{B}}\tau}\right)\right)\right)}e^{-p_{e}\kappa_{2}\left(\frac{\cos(\omega_{\mathrm{B}}\tau)+3\omega_{\mathrm{B}}\tau\sin(\omega_{\mathrm{B}}\tau)-1}{\omega_{\mathrm{B}}^{2/3}}-9\tau^{\frac{2}{3}}C_{3,3}\left(\sqrt[3]{\omega_{\mathrm{B}}\tau}\right)\right)}\\
\end{eqnarray}
\end{widetext}
in one and two dimensions. For three dimensions where the coherence is described by
\begin{eqnarray}
\label{eq:10}
\nonumber
\tilde{G}^{\mathrm{3D}}_{\mathrm{DD}}(\tau)
&\approx&  \alpha(\tau)\left( 1 + i\frac{p_{e}\kappa_{3}\tau}{N_{0}} \right. \\
& & \left. - \frac{p_{e}\kappa_{3}}{N_{0}}\tau (i(\gamma + \ln(\kappa_{3}\tau/N_{0})) + \pi/2)\right)^{N_{0}},\;\;\;\;
\end{eqnarray}
the limit is more difficult to obtain due to the $\ln(\kappa_{3}\tau/N_{0})$ term in the phase component. By only considering the amplitude we can at least obtain the decay relation in the limit $N_{0}\rightarrow \infty$, which is given by
\begin{eqnarray}
\label{eq:10axx}
|G^{\mathrm{3D}}_{\mathrm{DD}}(\tau)| &=& e^{-\frac{\pi}{2}p_{e}\kappa_{3}\tau}
\end{eqnarray}
and shows a simple exponential decay.

With the solutions of the integrals in Eq.~\eqref{eq:1aa} and Eq.~\eqref{eq:2} for the dipole-dipole and van der Waals potentials,  respectively, we can find a solution for a hybrid potential of these two defined by
\begin{eqnarray}
\label{eq:14}
U(r) = \left \{ \begin{array}{c} \frac{C_{3}}{r^{3}} \;\; r<r_{1} \\ \frac{C_{6}}{r^{6}} \;\; r \geq r_{1} \end{array} \right \},
\end{eqnarray}
with $\omega_{1} =  \frac{C_{3}}{r_{1}^{3}} = \frac{C_{6}}{r_{1}^{6}}$.
Here, we restrict ourselves to the three-dimensional case. The formulation for the function $\gamma_{\mathrm{Hyb}}(\tau)$ is given by
\begin{widetext}
\begin{eqnarray}
\label{eq:15}
\gamma_{\mathrm{Hyb}}(\tau) &=& \left(\frac{3}{r_{0}^{3}-r_{\mathrm{B}}^{3}}\right)\left( \int_{r_{\mathrm{B}}}^{r_{1}}dr r^{2} e^{i(C_{3}/r^{3})\tau} + \int_{r_{1}}^{r_{0}}dr r^{2} e^{i(C_{6}/r^{6})\tau}\right)\\
&=&\left (\frac{\omega_{\mathrm{B}}e^{i\omega_{1}\tau}-\omega_{1}e^{i\omega_{\mathrm{B}}\tau}}{\omega_{1}(\omega_{\mathrm{B}}-\omega_{0})}\right)\omega_{0}+\left (\frac{\omega_{\mathrm{B}}\omega_{0}\tau}{\omega_{\mathrm{B}}-\omega_{0}} \right) (i\left(\mathrm{Ci}(\omega_{\mathrm{B}}\tau) - \mathrm{Ci}(\omega_{1}\tau) \right)  - \left(\mathrm{Si}(\omega_{\mathrm{B}}\tau)-\mathrm{Si}(\omega_{1}\tau)\right))\\\nonumber
& & + \left (\frac{\sqrt{\tilde{\omega}_{1}}e^{i\tilde{\omega}_{0}\tau}-\sqrt{\tilde{\omega}_{0}}e^{i\tilde{\omega}_{1}\tau}}{\sqrt{\tilde{\omega}_{1}}(\sqrt{\tilde{\omega}}_{\mathrm{B}}-\sqrt{\tilde{\omega}}_{0})}\right)\sqrt{\tilde{\omega}_{\mathrm{B}}} + \dots\\\nonumber
& & + \left (\frac{2\sqrt{\tilde{\omega}_{\mathrm{B}}\tilde{\omega}_{0}\tau}}{\sqrt{\tilde{\omega}_{\mathrm{B}}}-\sqrt{\tilde{\omega}_{0}}} \right) (i\left(C_{0,2}(\sqrt{\tilde{\omega}_{1}\tau}) - C_{0,2}(\sqrt{\tilde{\omega}_{0}\tau}) \right)  - \left(S_{0,2}(\sqrt{\tilde{\omega}_{1}\tau})-S_{0,2}(\sqrt{\tilde{\omega}_{0}\tau})\right)),
\end{eqnarray}
\end{widetext}
where $\omega_{0} = \kappa_{3}\frac{1}{N_{0}}$ and $\sqrt{\tilde{\omega}_{0}} = \sigma_{3}\frac{1}{N_{0}}$. Since $\omega_{1} = \frac{C_{3}}{r^{3}_{1}} = \frac{C_{6}}{r^{6}_{1}} = \tilde{\omega}_{1}$ we find $\sqrt{\tilde{\omega}_{\mathrm{B}}} = \frac{\omega_{\mathrm{B}}}{\sqrt{\omega_{1}}}$ and $\sqrt{\tilde{\omega}_{0}} = \frac{\omega_{0}}{\sqrt{\omega_{1}}}$, which give us the relation $\frac{\sqrt{\tilde{\omega}_{0}}}{\sqrt{\omega_{1}}} = \frac{\omega_{0}}{\omega_{1}}$ that is equal to $\frac{\kappa_{3}}{\omega_{1}}=\frac{\sigma_{3}}{\sqrt{\omega_{1}}}$. Using these relations we obtain for
\begin{eqnarray}
\label{eq:16}
\nonumber
& & \left (\frac{\omega_{\mathrm{B}}e^{i\omega_{1}\tau}-\omega_{1}e^{i\omega_{\mathrm{B}}\tau}}{\omega_{1}(\omega_{\mathrm{B}}-\omega_{0})}\right)\omega_{0} + \left (\frac{\sqrt{\tilde{\omega}_{1}}e^{i\tilde{\omega}_{0}\tau}-\sqrt{\tilde{\omega}_{0}}e^{i\tilde{\omega}_{1}\tau}}{\sqrt{\tilde{\omega}_{1}}(\sqrt{\tilde{\omega}}_{\mathrm{B}}-\sqrt{\tilde{\omega}}_{0})}\right)\sqrt{\tilde{\omega}_{\mathrm{B}}} \\\nonumber
& & = e^{i\frac{\omega^{2}_{0}}{\omega_{1}}\tau}\left(1 + \frac{\omega_{0}\left(1 - e^{i\left(\omega_{\mathrm{B}}-\frac{\omega^{2}_{0}}{\omega_{1}} \right)\tau} \right)}{\omega_{\mathrm{B}}-\omega_{0}} \right)\\
& &\approx  \left(1 + \frac{\kappa_{3}\left(1  - e^{i\omega_{\mathrm{B}}\tau}\right)}{N_{0}\omega_{\mathrm{B}}} \right).
\end{eqnarray}
With the relations of Eq.~\eqref{eq:8} and Eq.~\eqref{eq:8a} we acquire
\begin{widetext}
\begin{eqnarray}
\label{eq:18}
\gamma_{\mathrm{Hyb}}(\tau) &=& \left(1 + \frac{\kappa_{3}\left(1  - \cos(\omega_{\mathrm{B}}\tau) + i\sin(\omega_{\mathrm{B}}\tau)\right)}{N_{0}\omega_{\mathrm{B}}}  \right. \\\nonumber
& & + \frac{\kappa_{3}}{N_{0}}(i\tau\left(\mathrm{Ci}(\omega_{\mathrm{B}}\tau) - \mathrm{Ci}(\omega_{1}\tau) \right)  - \tau\left(\mathrm{Si}(\omega_{\mathrm{B}}\tau)-\mathrm{Si}(\omega_{1}\tau)\right))\\\nonumber
& &\left. + \frac{2\sigma_{3}}{N_{0}}\left(i\sqrt{\tau}\left(C_{0,2}(\sqrt{\tilde{\omega}_{1}\tau}) - C_{0,2}(\sqrt{\tilde{\omega}_{0}\tau}) \right)  - \sqrt{\tau}\left(S_{0,2}(\sqrt{\tilde{\omega}_{1}\tau})-S_{0,2}(\sqrt{\tilde{\omega}_{0}\tau})\right)\right) \right).
\end{eqnarray}
\end{widetext}
For $\omega_{\mathrm{B}} \rightarrow \infty$ the resultant expression is
\begin{eqnarray}
\label{eq:19}
\nonumber
\gamma_{\mathrm{Hyb}}(\tau) &=& 1 - i\frac{\kappa_{3}\tau}{N_{0}}\mathrm{Ci}(\omega_{1}\tau) - \frac{\kappa_{3}\tau}{N_{0}}\left( \frac{\pi}{2}-\mathrm{Si}(\omega_{1}\tau)\right)\\
& & + \frac{2\sigma_{3}}{N_{0}}\sqrt{\tau}(iC_{0,2}(\sqrt{\omega_{1}\tau})-S_{0,2}(\sqrt{\omega_{1}\tau})).
\end{eqnarray}
For the function $\tilde{G}_{\mathrm{Hyb}}(\tau)$ we can write
\begin{widetext}
\begin{eqnarray}
\label{eq:19}
 \tilde{G}_{\mathrm{Hyb}}(\tau) &\approx&  \alpha(\tau)\left(1 - i\frac{p_{e}\kappa_{3}\tau}{N_{0}}\mathrm{Ci}(\omega_{1}\tau) - \frac{p_{e}\kappa_{3}\tau}{N_{0}}\left( \frac{\pi}{2}-\mathrm{Si}(\omega_{1}\tau)\right) + \frac{2p_{e}\sigma_{3}}{N_{0}}\sqrt{\tau}(iC_{0,2}(\sqrt{\omega_{1}\tau})-S_{0,2}(\sqrt{\omega_{1}\tau})) \right)^{N_{0}}\\
&\approx&  e^{i\left(\omega\tau + \phi-p_{e}\kappa_{3}\tau\mathrm{Ci}(\omega_{1}\tau) + 2p_{e}\sigma_{3}\sqrt{\tau}C_{0,2}(\sqrt{\omega_{1}\tau})\right)}e^{-p_{e}\kappa_{3}\tau\left( \frac{\pi}{2}-\mathrm{Si}(\omega_{1}\tau)\right) - 2p_{e}\sigma_{3}\sqrt{\tau}S_{0,2}(\sqrt{\omega_{1}\tau})}
\end{eqnarray}
\end{widetext}
and the corresponding Ramsey signal is given by
\begin{eqnarray}
\label{eq:20}
\nonumber
P(n,\tau) &=& 2p_{g}p_{e}\Re\left\{ 1+ e^{i\left(-p_{e}\kappa_{3}\tau\mathrm{Ci}(\omega_{1}\tau) + 2p_{e}\sigma_{3}\sqrt{\tau}C_{0,2}(\sqrt{\omega_{1}\tau})\right)} \right. \\\nonumber
& & \left. \times \alpha(\tau)e^{-p_{e}\kappa_{3}\tau\left( \frac{\pi}{2}-\mathrm{Si}(\omega_{1}\tau)\right) - 2p_{e}\sigma_{3}\sqrt{\tau}S_{0,2}(\sqrt{\omega_{1}\tau})} \right\},
\end{eqnarray}
where $-p_{e}\kappa_{3}\tau\mathrm{Ci}(\omega_{1}\tau) + 2p_{e}\sigma_{3}\sqrt{\tau}C_{0,2}(\sqrt{\omega_{1}\tau})$ is the dynamical phase shift, while the decay is described by $e^{-p_{e}\kappa_{3}\tau\left( \frac{\pi}{2}-\mathrm{Si}(\omega_{1}\tau)\right)- 2p_{e}\sigma_{3}\sqrt{\tau}S_{0,2}(\sqrt{\omega_{1}\tau})}$.

\section{Appendix D: Anisotropy}
The treatment presented in the previous section is not limited to an description of isotropic interactions between particles. Anisotropic interactions of the kind
\begin{eqnarray}
\label{Aneq:1}
U(r,\theta,\phi) &=& \left \{ \begin{array}{c} \frac{C_{3}f(\theta,\phi)}{r^{3}}\;\; \mathrm{DD} \\ \frac{C_{6}\hat{f}(\theta,\phi)}{r^{6}}\;\; \mathrm{vdW} \end{array} \right.
\end{eqnarray}
result in similar expressions as derived in the previous section. For an anisotropic van der Waals interaction $U(r,\theta,\phi) = C_{6}\hat{f}(\theta,\phi)/r^{6}$ we obtain
\begin{widetext}
\begin{eqnarray}
\label{Aneq:2}
\gamma^{1\mathrm{D}}_{\mathrm{vdW}}(\tau) &\approx& 1 + \frac{\sigma_{1}\left(1 - \exp(i\omega_{\mathrm{B}}\tau_{\theta,\phi})\right)}{N_{0}\sqrt[6]{\omega_{\mathrm{B}}}} + \frac{6\sigma_{1}}{N_{0}}\sqrt[6]{\tau_{\theta,\phi}}\left( i\left(\tilde{C}_{1}(\sqrt[6]{\omega_{\mathrm{B}}\tau_{\theta,\phi}}) \right)  - \left(\tilde{S}_{1}(\sqrt[6]{\omega_{\mathrm{B}}\tau_{\theta,\phi}})\right)\right),\\
\gamma^{2\mathrm{D}}_{\mathrm{vdW}}(\tau) &\approx& 1 + \frac{1}{2\pi}\int_{0}^{2\pi}d\phi \left(\frac{\sigma_{2}\left(1 - \exp(i\omega_{\mathrm{B}}\tau_{\theta,\phi})\right)}{N_{0}\sqrt[3]{\omega_{\mathrm{B}}}} + \frac{3\sigma_{2}}{N_{0}}\sqrt[3]{\tau_{\theta,\phi}}\left( i\left(\tilde{C}_{2}(\sqrt[3]{\omega_{\mathrm{B}}\tau_{\theta,\phi}}) \right)  - \left(\tilde{S}_{2}(\sqrt[3]{\omega_{\mathrm{B}}\tau_{\theta,\phi}})\right)\right)\right),\\\nonumber
\gamma^{3\mathrm{D}}_{\mathrm{vdW}}(\tau) &\approx& 1 + \frac{1}{4\pi}\int_{0}^{2\pi}\int_{-1}^{1}d\phi d(\cos(\theta))\left(\frac{\sigma_{3}\left(1 - \exp(i\omega_{\mathrm{B}}\tau_{\theta,\phi})\right)}{N_{0}\sqrt{\omega_{\mathrm{B}}}} + \frac{2\sigma_{3}}{N_{0}}\sqrt{\tau_{\theta,\phi}}\left( i\left(\tilde{C}_{3}(\sqrt{\omega_{\mathrm{B}}\tau_{\theta,\phi}}) \right)  - \left(\tilde{S}_{3}(\sqrt{\omega_{\mathrm{B}}\tau_{\theta,\phi}})\right)\right)\right),\\
\end{eqnarray}
\end{widetext}
where $\tau_{\theta,\phi} = \hat{f}(\theta,\phi)\tau$. The results for the dipole-dipole interaction and in the case of the hybrid model are obtained analogously by subtituting $\tau$ with $\tau_{\theta,\phi}$ and by leaving the angular integration untouched.\\
\indent For example in the case of a three dimensional ensemble with a homogeneous atom distribution where the interaction is given by an anisotropic van der Waals interaction, where $\hat{f}(\theta,\phi) = (1-3\cos^{2}(\theta))^{2}$ \cite{Takei2015}, and where we take $\omega_{\mathrm{B}} \rightarrow \infty$, we obtain
\begin{eqnarray}
\label{Aneq:3}
\gamma^{\mathrm{3D}}_{\mathrm{vdW}} &\approx& 1 + \frac{\sigma_{3}}{N_{0}}\sqrt{\frac{\pi}{8}}\sqrt{\tau}(i -1)\int_{-1}^{1}d(\cos(\theta))\sqrt{\hat{f}(\theta)}\;\;\;\\
&=& 1 + \frac{8\sigma_{3}}{3\sqrt{3}N_{0}}\sqrt{\frac{\pi}{8}}\sqrt{\tau}(i -1),
\end{eqnarray}
where in contrast to a factor of two in the isotropic case (Eq.~\eqref{eq:9bx}) we get a factor of $8/(3\sqrt{3})$.
\section{Appendix E: Correlations}
To obtain further information of the state after the Ramsey process different projection operations on eigenstates need to be performed on the system. This could be realized in an experiment with single site resolution where the population content in an eigenstate of a particular atom can be easily discriminated from all the other atoms. For simplicity and without loss of generality we set the phase $\phi$ of the pumping matrix $A$ to zero. Also for more clarity the frequency $\omega$ is accompanied by the index of the atom of origin and written as $\omega_{j}$ for atom $j$.

For example, the signal for measuring two atoms in the excited state is given by
\begin{widetext}
\begin{eqnarray}
\label{Eq.10}
\nonumber
P_{jk}^{(N)}(\tau) &=& \langle \Psi(\tau)|e\rangle_{j}\langle e|_{j}\otimes |e\rangle_{k}\langle e|_{k} |\Psi(\tau)\rangle_{N}\\
&=& 2p_{g}^{2}p_{e}^{2}\Re \left[ 2 + e^{i(\omega_{j}-\omega_{k})\tau}\prod_{\begin{smallmatrix} l=1 \\ l \neq j \neq k\end{smallmatrix}}^{N}\left( p_{g} + p_{e}e^{i(U_{jl}-U_{kl})\tau}\right)+ e^{i\omega_{j}\tau}\left( 1 + e^{iU_{jk}\tau}\right)\prod_{\begin{smallmatrix} l=1 \\ l \neq j \neq k\end{smallmatrix}}^{N}\left( p_{g} + p_{e}e^{iU_{jl}\tau}\right) \right.\\\nonumber
& & \left.+ e^{i\omega_{k}\tau} \left( 1 + e^{iU_{jk}\tau}\right)\prod_{\begin{smallmatrix} l=1 \\ l \neq j \neq k\end{smallmatrix}}^{N}\left( p_{g} + p_{e}e^{iU_{kl}\tau}\right) +  \left(e^{i\left(\omega_{j}+\omega_{k}+ U_{jk}\right)\tau}\right)\prod_{\begin{smallmatrix} l=1 \\ l \neq j \neq k\end{smallmatrix}}^{N}\left( p_{g} + p_{e}e^{i(U_{jl}+U_{kl})\tau}\right)\right].
\end{eqnarray}
\end{widetext}
The expression in Eq.~\eqref{Eq.10} can be used to obtain the signal for two simultaneously measured spins, which is described by
$\langle \Psi(\tau)|\hat{\sigma}_{z}^{(j)}\otimes\hat{\sigma}_{z}^{(k)}| \Psi(\tau)\rangle_{N} = 2^{2}P_{jk}^{(N)}(\tau) - 2P_{j}^{(N)}(\tau) - 2P_{k}^{(N)}(\tau) +1,$ where $\hat{\sigma}_{z}^{(j)} = |e\rangle_{j}\langle e|_{j} - |g\rangle_{j}\langle g|_{j}$.
This allows us to derive the spin-spin correlation dependence in an $N$-particle ensemble given by
\begin{widetext}
 \begin{eqnarray}
\label{Eq.11a}
C_{jk}(\tau) &=& \langle \Psi(\tau)|\hat{\sigma}_{z}^{(j)}\otimes\hat{\sigma}_{z}^{(k)}| \Psi(\tau)\rangle_{N} - \langle \Psi(\tau)|\hat{\sigma}_{z}^{(j)}| \Psi(\tau)\rangle_{N} \langle \Psi(\tau)| \hat{\sigma}_{z}^{(k)}| \Psi(\tau)\rangle_{N}\\\nonumber
&=& 2^{2}\left( P_{jk}^{(N)} - P_{j}^{(N)}P_{k}^{(N)}\right),
\end{eqnarray}
\end{widetext}
which in the case where all interactions are set to zero results in $C_{jk}(\tau) = 0$ \cite{Feig2013,Hazzard2014,Zeiher2016}.\\

The results for $|e\rangle_{j}\langle e|_{j}\otimes |g\rangle_{k}\langle g|_{k}$ as well as $|g\rangle_{j}\langle g|_{j}\otimes |e\rangle_{k}\langle e|_{k}$ and $|g\rangle_{j}\langle g|_{j}\otimes |g\rangle_{k}\langle g|_{k} $ can be constructed analogously.\\
\indent Setting all interactions to zero results in
$P_{jk}^{(N)}(\tau) = 2p_{g}^{2}p_{e}^{2}\left[ 2 + \cos((\omega_{j}-\omega_{k})\tau) +  2\cos(\omega_{j}\tau) \right.$  $\left.+ 2\cos(\omega_{k}\tau) + \cos((\omega_{j}+\omega_{k})\tau) \right],$
which for $\omega_{j} = \omega_{k} = \omega$ leads to $ 2p_{g}^{2}p_{e}^{2}\left[ 3 + 4\cos(\omega\tau) + \cos(2\omega\tau) \right]$.\\
\indent This can be generalized further on to the results to simultaneously measure $m$ atoms out of an ensemble of $N$ in the excited state which is given by
\begin{widetext}
\begin{eqnarray}
\label{Eq.13}
P_{j_{1},\dots,j_{m}}^{(N)}(\tau) &=& \langle \Psi(\tau)|e\rangle_{j_{1}}\langle e|_{j_{1}}\otimes \dots \otimes |e\rangle_{j_{m}}\langle e|_{j_{m}}| \Psi(\tau)\rangle_{N}\\
&=& 2p_{g}^{m}p_{e}^{m}\Re \left[2^{m-1} +  \sum_{\begin{smallmatrix} I \subset \{j_{1},\dots,j_{m}\} \\ |I| > 0 \end{smallmatrix}} \sum_{\begin{smallmatrix} \lambda \in \Lambda_{I} \end{smallmatrix}} \prod_{a \in I} e^{i\lambda_{a}\omega_{a}\tau}\prod_{\begin{smallmatrix}J = \{b,c\} \\ J \subset I \end{smallmatrix}} e^{i(\lambda_{b}+\lambda_{c})\frac{U_{bc}}{2}\tau} \right. \\\nonumber
& & \left. \times \prod_{d \in \{j_{1},\dots,j_{m}\} \backslash I} \left(1 + e^{i\sum_{f\in I} \lambda_{f}U_{fd}\tau}  \right)\prod_{h\in \{1,\dots,N\} \backslash \{j_{1},\dots,j_{m}\}} \left(p_{g} + p_{e} e^{i\sum_{k\in I} \lambda_{k}U_{kh}\tau} \right) \right],
\end{eqnarray}
\end{widetext}
where $\Lambda_{I} = \{ \lambda = (\lambda_{k_{1}},\dots,\lambda_{k_{|I|}}) | k_{1} < \dots < k_{|I|} \wedge \lambda \in \{1\}\times\{-1,1\}^{\times |I|-1}\}$.
The corresponding signal for $m$ simultaneously measured spins is defined by
\begin{widetext}
\begin{eqnarray}
\label{Eq.14}
\langle \Psi(\tau)|\hat{\sigma}_{z}^{(j_{1})}\otimes\dots \otimes \hat{\sigma}_{z}^{(j_{m})}| \Psi(\tau)\rangle_{N} = \sum_{k=0}^{m} (-1)^{k}2^{m-k}\sum_{\begin{smallmatrix} I \subset \{j_{1},\dots,j_{m}\} \\ |I| = m- k \end{smallmatrix}} P_{I}(\tau).
\end{eqnarray}
\end{widetext}
Analogous to the procedure in Eq.~\eqref{Eq.11a}, the correlation function for $m$ spins can be obtained from these measurements.\\
\indent Again, setting all interaction terms to zero agrees well with the general result of the Ramsey signal for measuring $m$ noninteracting atoms simultaneously in the excited state with $\omega_{a} = \omega$, which is
\begin{eqnarray}
\label{Eq.15}
\nonumber
P_{m}(\tau) = 2p_{g}^{m}p_{e}^{m}\left[ \left( \begin{array}{c} 2m-1 \\ m \end{array} \right) + \sum_{k=1}^{m}\left( \begin{array}{c} 2m \\ m+k \end{array} \right)\cos(k\omega\tau)\right].\\
\end{eqnarray}
Furthermore, it is not difficult to derive the Ramsey signal for initial states that are different than $|g\rangle^{\otimes N}$. Choosing from the separable many-particle states described by $|\Psi\rangle_{I} = \bigotimes_{j\in I} |g\rangle_{j} \bigotimes_{k \in \tilde{N}\backslash I} |e\rangle_{k}$ for the initial state, where $I \subset \{1,\dots, N\} = \tilde{N}$, we obtain for
\begin{eqnarray}
\label{Eq.16}
|\Psi(\tau)\rangle_{N:I} = A^{\otimes N}\exp\left(-i\hat{H}\tau \right)A^{\otimes N}|\Psi\rangle_{I}
\end{eqnarray}
 in the case of $|e\rangle_{j}\langle e|_{j}$ and $j \in I$
 \begin{widetext}
\begin{eqnarray}
\label{Eq.17}
P_{j:I}^{(N)}(\tau) =  2p_{g}p_{e}\Re\left[ 1+ e^{i\omega_{j}\tau}\prod_{\begin{smallmatrix} k\in I \\ k \neq j \end{smallmatrix}}\left( p_{g} + p_{e}e^{iU_{jk}\tau}\right)\prod_{k\in \tilde{N}\backslash I }\left( p_{g}e^{iU_{jk}\tau} + p_{e}\right) \right],
\end{eqnarray}
\end{widetext}
while for $j \in \tilde{N}\backslash I$ we derive
\begin{widetext}
\begin{eqnarray}
\label{Eq.18}
P_{j:I}^{(N)}(\tau) = 1 - 2p_{g}p_{e}\Re\left[ 1+ e^{i\omega_{j}\tau}\prod_{k\in I }\left( p_{g} + p_{e}e^{iU_{jk}\tau}\right)\prod_{\begin{smallmatrix} k\in \tilde{N}\backslash I\\ k \neq j  \end{smallmatrix}}\left( p_{g}e^{iU_{jk}\tau} + p_{e}\right) \right].
\end{eqnarray}
\end{widetext}
This leads to more complex expressions (similar to Eq.~\eqref{Eq.14}) for higher evaluation operators like $|e\rangle_{j_{1}}\langle e|_{j_{1}} \otimes |e\rangle_{j_{2}}\langle e|_{j_{2}}$  if the readout state and the initial state have coinciding state components as given in Eq.~\eqref{Eq.18}. To overcome this problem it is best to use only complementary state components with respect to the initial state for the readout operation. For example, this was already done in the initial discussion to derive the result in Eq.~\eqref{Eq.13}, where the initial state is $|g\rangle^{\otimes N}$ and the readout operator with $|e\rangle_{j_{1}}\langle e|_{j_{1}}\otimes \dots \otimes |e\rangle_{j_{m}}\langle e|_{j_{m}}|$ contains no ground-state state components. For example, for $|\Psi \rangle_{I}$ with $I \subset \tilde{N}$ we derive for a two-particle readout operator $|e\rangle_{j_{1}}\langle e|_{j_{1}} \otimes |g\rangle_{j_{2}}\langle g|_{j_{2}}$ with $j_{1} \in I$ and $j_{2} \in \tilde{N}\backslash I$ the analogous formulation to Eq.~\eqref{Eq.10} with
\begin{widetext}
\begin{eqnarray}
\label{Eq.19}
Q_{j_{1}j_{2}:I}^{(N)}(\tau) &=& \langle \Psi(\tau)|e\rangle_{j_{1}}\langle e|_{j_{1}}\otimes |g\rangle_{j_{2}}\langle g|_{j_{2}} |\Psi(\tau)\rangle_{N:I}\\\nonumber
&=& 2p_{g}^{2}p_{e}^{2}\Re \Bigg[ 2 + e^{i(\omega_{j_{1}}-\omega_{j_{2}})\tau}\prod_{\begin{smallmatrix} l\in I \\ l \neq j_{1} \end{smallmatrix}} \left( p_{g} + p_{e}e^{i\left(U_{j_{1}l}-U_{j_{2}l}\right)\tau}\right)\prod_{\begin{smallmatrix} l\in \tilde{N}\backslash I \\ l \neq j_{2} \end{smallmatrix}} \left( p_{g}e^{i\left(U_{j_{1}l}-U_{j_{2}l}\right)\tau} + p_{e}\right)\\\nonumber
& &  + e^{i\omega_{j_{1}}\tau}\left( 1 + e^{iU_{j_{1}j_{2}}\tau}\right)\prod_{\begin{smallmatrix} l\in I \\ l \neq j_{1} \end{smallmatrix}}\left( p_{g} + p_{e}e^{iU_{j_{1}l}\tau}\right)\prod_{\begin{smallmatrix} l\in \tilde{N}\backslash I \\ l \neq j_{2} \end{smallmatrix}}\left( p_{g}e^{iU_{j_{1}l}\tau} + p_{e}\right)\\\nonumber
& &  + e^{i\omega_{j_{2}}\tau} \left( 1 + e^{iU_{j_{1}j_{2}}\tau}\right)\prod_{\begin{smallmatrix} l\in I \\ l \neq j_{1} \end{smallmatrix}}\left( p_{g} + p_{e}e^{iU_{j_{2}l}\tau}\right)\prod_{\begin{smallmatrix} l\in \tilde{N}\backslash I \\ l \neq j_{2} \end{smallmatrix}}\left( p_{g}e^{iU_{j_{2}l}\tau} + p_{e}\right)\\\nonumber
& & \left. +  \left(e^{i\left(\omega_{j_{1}}+\omega_{j_{2}}+ U_{j_{1}j_{2}}\right)\tau}\right)\prod_{\begin{smallmatrix} l\in I \\ l \neq j_{1} \end{smallmatrix}}\left( p_{g} + p_{e}e^{i\left( U_{j_{1}l}+U_{j_{2}l}\right)\tau}\right)\prod_{\begin{smallmatrix} l\in \tilde{N} \backslash I \\ l \neq j_{2} \end{smallmatrix}}\left( p_{g}e^{i\left( U_{j_{1}l}+U_{j_{2}l}\right)\tau} + p_{e}\right)\right].
\end{eqnarray}
\end{widetext}
Using the identity $|e\rangle_{j_{k}}\langle e|_{j_{k}} +  |g\rangle_{j_{k}}\langle g|_{j_{k}} = 1$ all other solutions to read-out operators, where maximally two particles are simultaneously measured in an ensemble of $N$, can be constructed. For example, for the spin measurement $\hat{\sigma}_{z}^{(j_{1})}\otimes\hat{\sigma}_{z}^{(j_{2})} = -2^{2}|e\rangle_{j_{1}}\langle e|_{j_{1}}\otimes |g\rangle_{j_{2}}\langle g|_{j_{2}} + 2|e\rangle_{j_{1}}\langle e|_{j_{1}} + 2|g\rangle_{j_{2}}\langle g|_{j_{2}} -1$ we obtain
\begin{widetext}
\begin{eqnarray}
\label{Eq.20}
\langle \Psi(\tau)|\hat{\sigma}_{z}^{(j_{1})}\otimes\hat{\sigma}_{z}^{(j_{2})}| \Psi(\tau)\rangle_{N:I} = -2^{2}Q_{j_{1}j_{2}:I}^{(N)}(\tau) + 2Q_{j_{1}:I}^{(N)}(\tau) + 2Q_{j_{2}:I}^{(N)}(\tau) -1,
\end{eqnarray}
\end{widetext}
where $Q_{j_{1}:I}^{(N)}(\tau) = P_{j_{1}:I}^{(N)}(\tau)$ since $j_{1}\in I$ and the sign is changed in comparison to the spin-spin average value derived from Eq.~\eqref{Eq.10}. This is caused by the particle $j_{2}$ initially being in the excited state. The Ramsey signal for many initially entangled states can be calculated by first solving the Ramsey signal for the separable many-particle basis and superposing the solutions with the corresponding coefficients. Nevertheless, for increasing $N$ this construction becomes an impossible task due to the exponentially growing basis set with respect to the particle number.\\

\end{document}